\documentclass[10pt,journal, final]{IEEEtran}
\usepackage{ifpdf}
\usepackage{cite}
\usepackage{algorithmic}
\usepackage{algorithm}

\ifCLASSINFOpdf
\usepackage[pdftex]{graphicx}

\else
 \usepackage[dvips]{graphicx}
\fi
\usepackage[cmex10]{amsmath}
\usepackage{epstopdf}
\usepackage{array}
\usepackage{flushend}
\usepackage{balance}
\usepackage{eqparbox}

\usepackage[tight,footnotesize]{subfigure}
\usepackage{fixltx2e}
\usepackage{color}
\usepackage{float}

\usepackage{dblfloatfix}
\usepackage{url}
\usepackage{cite}
\usepackage{subfigure}
\usepackage{amsmath}%[centertags][tbtags][sumlimits][nosumlimits][intlimits][namelimits][nonamelimits]
\usepackage{epstopdf}

\newtheorem{remark}{Remark}
\newtheorem{lemma}{Lemma}

\usepackage{amssymb}
\usepackage{color}

\usepackage{cuted}

\begin{document}
\allowdisplaybreaks[3]
\title{Three-Dimensional Spatial Correlation Modeling for Cylindrical mMIMO Arrays in HAPS}
\setlength{\columnsep}{0.21 in}

\author{Shasha~Liu,~\IEEEmembership{Student Member,~IEEE,}
Abla~Kammoun,~\IEEEmembership{Member,~IEEE,}      
        and
        Mohamed-Slim~Alouini,~\IEEEmembership{Fellow,~IEEE}

\thanks {
Shasha Liu, Abla~Kammoun, and Mohamed-Slim Alouini are with the Division of Computer, Electrical and Mathematical Sciences and Engineering, King Abdullah University of Science and Technology, Thuwal 23955-6900, Saudi Arabia (e-mail: shasha.liu@kaust.edu.sa, abla.kammoun@kaust.edu.sa, slim.alouini@kaust.edu.sa).

%Hayssam Dahrouj is with the department of Electrical Engineering, University of Sharjah, Sharjah, UAE (e-mail: hayssam.dahrouj@gmail.com).
}
}
\maketitle
\begin{abstract}
High-altitude platform stations (HAPS) are envisioned as a key component of future wireless networks, enabling ultra-wide coverage and providing direct connectivity to users with cylindrical massive multiple-input multiple-output (mMIMO) systems. Exploiting the channel degrees of freedom necessitates accurate modeling and characterization of three-dimensional (3D) channels in the presence of spatial correlation functions (SCFs).
However, existing spatial correlation models are primarily developed for planar or linear antenna arrays and cannot be directly applied to cylindrical geometries commonly adopted by HAPS platforms. To address this limitation, this paper derives an exact closed-form expression for the SCF of 3D MIMO channels with antenna elements arranged in a cylindrical array. The proposed formulation is based on the spherical harmonic expansion (SHE) of plane waves and accommodates arbitrary antenna radiation patterns and angular distributions through the Fourier series (FS) coefficients of the power azimuth and zenith spectra. The derived SCF is validated through Monte Carlo simulations under standard-compliant settings.
\end{abstract}
\begin{IEEEkeywords}
Massive MIMO, correlation matrix, HAPS, cylindrical arrays.
\end{IEEEkeywords}
\IEEEpeerreviewmaketitle
\section{Introduction}
High-altitude platform stations (HAPS) have emerged as a promising enabling technology for beyond-5G and sixth-generation (6G) wireless networks, as they are capable of providing wide-area coverage while overcoming many limitations of terrestrial infrastructure \cite{mohammed2011role, shibata2020system}. Operating at altitudes of approximately $20$ km, solar-powered HAPS can directly deliver Internet connectivity to user equipment (UEs), making them particularly suitable for serving hard-to-access areas such as mountainous regions, oceans, and glaciers. Compared with conventional terrestrial base stations, a single HAPS can cover a substantially larger geographical area, and, relative to satellite systems, HAPS can offer communication services with lower deployment cost and reduced latency. 
\par
% claim the bennifit og MIMO 
Due to the increasing scarcity of wireless spectrum resources, massive multiple-input multiple-output (mMIMO) has emerged as a key enabling technology for 6G and beyond \cite{rusek2012scaling,larsson2014massive,hoydis2013massive}. By exploiting abundant spatial degrees of freedom, mMIMO provides substantial diversity, interference mitigation, and spatial multiplexing gains, thereby significantly improving spectral efficiency and network capacity. Recently, mMIMO-enabled beamforming has been widely investigated in non-terrestrial networks. For example, \cite{lin2021supporting} incorporated RSMA into satellite–UAV integrated networks to enhance interference management and spectral efficiency. In \cite{lin2022refracting}, active beamforming at a multi-antenna satellite and relay was jointly optimized with RIS phase shifts in a hybrid satellite–terrestrial relay network. The authors in \cite{zhi2024self} further considered a self-powered absorptive RIS-assisted satellite–terrestrial integrated network, where MIMO beamforming and RIS reflection coefficients were jointly designed to improve physical-layer security. \cite{an2024exploiting} proposed a multi-layer refracting RIS-assisted receiver architecture for HAP-simultaneous
wireless information and power transfer (SWIPT) networks, where multiple stacked RIS layers are deployed at the receiver side to enhance both information transmission and wireless energy harvesting.
\par
One of the key objectives of HAPS-based communication systems is to simultaneously enhance coverage and capacity. Owing to their large physical surface and payload capabilities, HAPS can support a large number of antenna elements, making them well suited for massive multiple-input multiple-output (mMIMO) deployments \cite{alam2021high,kurt2021vision}. By exploiting large-scale antenna arrays, mMIMO systems enable multiuser MIMO (MU-MIMO) transmission, whereby narrow beams can be dynamically steered toward individual UEs. This capability significantly improves spectral efficiency and overall system capacity. As a result, the configuration and geometry of the antenna array play a critical role in determining the performance of HAPS-based mMIMO systems.

Most existing studies on mMIMO systems for HAPS assume the use of linear or planar antenna arrays mounted on the underside of the platform, with all antenna elements oriented toward the Earth’s surface. While such configurations simplify system design and analysis, they inherently limit the coverage radius to approximately $20 $–$60$ km.  To address this limitation, Tashiro \emph{et al.} \cite{tashiro2021cylindrical} proposed a cylindrical massive MIMO architecture for HAPS communications, where a cylindrical antenna array was employed to overcome the coverage limitations of conventional planar arrays. Simulation results in \cite{tashiro2021cylindrical} demonstrated that the proposed architecture can achieve up to 10 dB SINR improvement at the coverage boundary and up to 2.1 times higher system capacity compared with planar-array-based massive MIMO systems. Building upon the concept of cylindrical arrays, Zhou \emph{et al.} \cite{zhou2023cylindrical} proposed a cylindrical array-based MIMO antenna system with a cell-fixation mechanism to maintain stable ground-cell coverage under HAPS mobility. To verify the feasibility of the proposed design, \cite{zhou2023cylindrical} developed and implemented a prototype cylindrical antenna system for a typical HAPS scenario with a UAV operating at an altitude of 20 km. Both MU-MIMO simulations and experimental measurements confirmed the effectiveness of the proposed cylindrical array architecture in achieving wide-area coverage and stable cell fixation. Riviello \emph{et al.} \cite{riviello2024multi} investigated multi-user multi-layer MIMO transmission with cylindrical arrays under the 3GPP 3D channel model for B5G/6G networks. By comparing cylindrical and conventional trisector planar arrays in realistic LOS, NLOS, and indoor scenarios, the authors demonstrated that cylindrical arrays provide more uniform cell coverage, mitigate sector-edge degradation, and achieve higher sum-rate and spectral efficiency, making them a promising antenna architecture for future massive MIMO systems. Motivated by these advantages, this paper adopts the cylindrical antenna array architecture to provide wide-area coverage, and improve beamforming performance.
\par
In addition, the substantial capacity gains promised by MIMO systems rely on the assumption that the individual subchannels are sufficiently uncorrelated. In practical deployments, however, channel fading often exhibits spatial correlation due to factors such as limited antenna spacing and constrained array geometries \cite{shiu2000fading, kang2006capacity}. 
In addition, 
realistic MIMO channels often exhibit significant spatial correlation due to the clustered nature of scattering in practical propagation environments \cite{shiu2000fading,chizhik2000effect}. Such spatial correlation generally degrades the performance of MIMO systems by reducing channel orthogonality, leading to losses in both capacity and error-rate performance \cite{gore2002statistical,forenza2007simplified}.
Consequently, accurately modeling spatial correlation is essential for realistic performance evaluation of HAPS-based mMIMO systems.

\par
The spatial correlation function (SCF) is a fundamental metric for characterizing the performance of multi-antenna systems, particularly in three-dimensional (3D) and full-dimension MIMO (FD-MIMO) networks.
Early works, such as \cite{michailidis2010three} derives the space–time correlation function for a $2\times 2$ HAPS-MIMO system using a 3D geometry-based single-scatterer model, accounting for factors such as platform zenith angle, array configuration, Doppler spread, and HAP displacement. However, this approach does not scale to mMIMO systems with hundreds of antenna elements and rich multi-scatterer environments, and would also incur prohibitively high computational complexity.
To overcome these limitations, the work in \cite{kammoun2015generalized} derived generalized SCFs for 3D MIMO channels by modeling individual antenna ports. This method avoids the aforementioned issues by the spherical harmonic expansion (SHE) of plane waves combined with trigonometric expansions of Legendre and associated Legendre polynomials, under prescribed angular distributions and antenna configurations. Such a framework enables a tractable analytical representation of spatial correlation in 3D propagation environments. Building upon this foundation, the work in  \cite{kammoun2016spatial} investigated SCFs for uniform circular arrays using a port-based model, while that in \cite{kammoun2017design} extended the analysis to FD-MIMO systems employing active antenna systems (AAS) with two-dimensional (2D) planar arrays. In the latter work, an element-wise modeling approach was adopted, which is particularly relevant for the design and optimization of 3D beamforming strategies.

\par
Despite these advances, HAPS systems often employ complex non-planar antenna architectures, such as cylindrical and hemispherical arrays \cite{abbasi2024hemispherical}, to achieve wide-area coverage. For such array geometries, the corresponding spatial correlation functions remain largely unexplored in the existing literature. Motivated by this gap, the primary objective of this paper is to derive an exact closed-form expression for the SCF of three-dimensional MIMO channels formed by individual antenna elements arranged in a cylindrical array. By exploiting the SHE of plane waves, the derived expression explicitly incorporates the effects of arbitrary angular distributions and antenna radiation patterns through the Fourier series (FS) coefficients of the power azimuth and zenith spectra. As a result, the proposed SCF formulation is applicable to general antenna patterns and arbitrary distributions of azimuth and zenith angles of departure (AoDs).
The parametric 3D channel model adopted in this work is inspired by standardized channel models, such as those specified by 3GPP \cite{3gpp_tr38811}. The second objective of this paper is to validate the proposed SCF through numerical simulations using antenna patterns and angular distributions defined in the standards. In particular, FS coefficients are computed and employed to obtain analytical correlation coefficients, which are shown to closely match Monte Carlo simulation results. 
\par
This paper is organized as follows. \ref{section: Antenna Array Configuration and 3D Channel Model} introduces the antenna configuration and element-level modeling. Section \ref{section: SCF} presents the derivation of SCFs for the antenna elements.  Section \ref{section: simulation} validates the derived SCFs through Monte Carlo simulations and investigates their impact on massive MIMO SINR performance. Finally, Section \ref{section: conclusion} concludes the paper.

\section{Antenna Array Configuration}
\label{section: Antenna Array Configuration and 3D Channel Model}
To achieve high-capacity and wide-area communication links in HAPS networks, the antenna system is a key enabler, as it directly affects both coverage and beamforming performance. In this section, we describe the antenna configuration and element modeling approach adopted by SoftBank \cite{tashiro2021cylindrical} for its HAPS architecture. We first discuss the spatial distribution of the antennas on the cylindrical surface and the bottom of the platform, followed by the characterization of each element’s radiation pattern and the corresponding array response formulation. This modeling framework forms the foundation for the subsequent correlation derivation.
\subsection{Antenna Configuration}
\begin{figure}[h]
\centering
\includegraphics[width=3in]{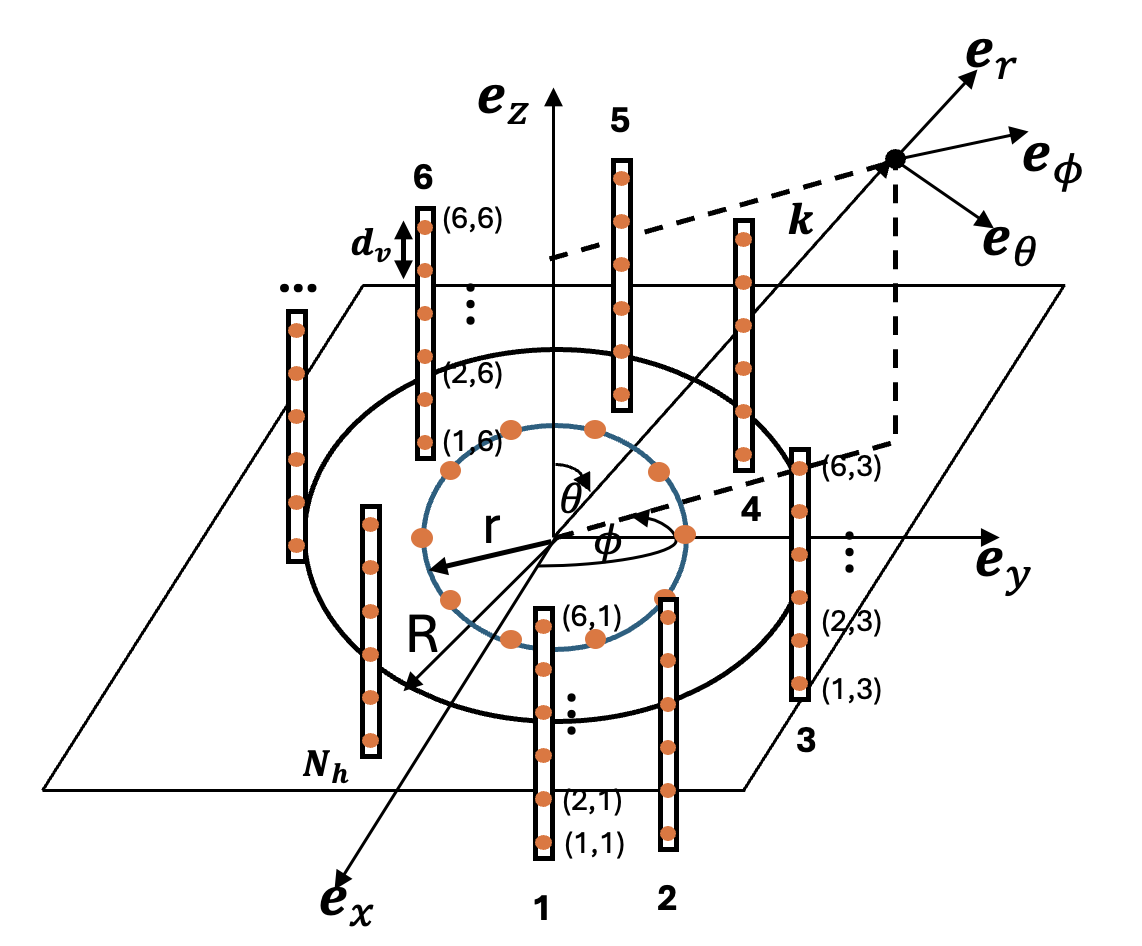}
\caption{Antenna configuration of HAPS}
\label{Fig: AN_Tx}
\end{figure}
%To improve the coverage of HAPS, reference \cite{tashiro2021cylindrical} proposes a large-scale cylinder-shaped array antenna. 
Fig. \ref{Fig: AN_Tx}
depicts the structure of the cylindrical antenna array adopted by Softbank. Specifically, in the horizontal direction,  $N_h$ antenna elements are arranged at regular intervals $d_h$ on the circumference of a circle of radius $R$. In the vertical direction, $N_v$ antenna elements are linearly placed at regular intervals $d_v$. In addition to $N_hN_v$ antenna elements on the curved surface of the cylindrical
array, some more elements are  placed on the bottom of the antenna in order to cover the area just below the aircraft. Thus, $N_b$ antenna elements are uniformly arranged at intervals of $d_b$ under the array antenna. Therefore, the total number of antenna elements can be expressed as $N_t=N_hN_v+N_b$. 
\par
\subsection{Antenna Element Modeling}
In this work, we adopt the antenna radiation patterns as defined in the 3GPP specifications, 
where the overall antenna gain is obtained as the combination of the zenith and azimuth patterns:
%The antenna element radiation pattern is given by 
\begin{equation}
    A_E(\phi, \theta)=G_{E,\text{max}}-\min\{-[A_{E,H}(\phi)+A_{E,V}(\theta)],A_{\text{m}}\} ,
\end{equation}
where,
\begin{equation}
A_{E, V}\left(\theta\right)=-\min \left\{12\left(\frac{\theta-90^{\circ}}{\theta_{3 \mathrm{dB}}}\right)^2, S L A_V\right\} \mathrm{dB},
\end{equation}
and 
\begin{equation}
A_{E, H}\left(\phi\right)=-\min \left\{12\left(\frac{\phi}{\phi_{3 \mathrm{dB}}}\right)^2, A_m\right\} \mathrm{dB}.
\end{equation}
Here, the azimuth angle $\phi$ and the zenith angle $\theta$ are defined with respect to the 
local spherical coordinate system of the antenna, which is oriented such that the 
boresight corresponds to $0^\circ$ azimuth and $90^\circ$ elevation, \footnote{%
The local coordinate system is constructed by taking the antenna axis as the $\mathbf{e}_z$ direction 
and deriving the local basis vectors $(\mathbf{e}_r, \mathbf{e}_\theta, \mathbf{e}_\phi)$ 
such that $\mathbf{e}_\phi$ is orthogonal to both $\mathbf{e}_r$ and $\mathbf{e}_z$.}
%In this expression, $\phi$ and $\theta$ denote the azimuth and elevation angles, respectively; 
whereas $A_{E,H}(\phi)$ and $A_{E,V}(\theta)$ represent the horizontal and vertical element radiation patterns, 
$G_{E,\text{max}}$ is the maximum directional gain of the antenna element, 
$\phi_{3\mathrm{dB}}$ and $\theta_{3\mathrm{dB}}$ are the half-power beamwidths (HPBW) in the azimuth and zenith planes, respectively, 
$SLA_{V}$ denotes the vertical side-lobe attenuation, and $A_m$ is the maximum attenuation.
The global field pattern of a vertically polarized antenna element in linear scale is $\sqrt{A_E(\phi, \theta)|_{\text{lin}}}=\sqrt{10^{\frac{G_{E,\text{max}}}{10}} g_E(\phi,\theta)}$, where $g_E(\phi,\theta) \approx g_{E,V}(\theta)g_{E,H}(\phi)$ with
\begin{equation}
\label{Eq:G_EV}
     g_{E,V}(\theta)=\max \left\{\exp{\left(-1.2\left(\frac{\theta-90^{\circ}}{\theta_{3 \mathrm{dB}}}\right)^2 \ln{10}\right)}, 10^{\frac{-SLA_V}{10}}\right\},
\end{equation}
\begin{equation}
\label{Eq:G_EH}
    g_{E,H}(\phi)=\max\left\{\exp{\left(-1.2\left(\frac{\phi}{\phi_{3 \mathrm{dB}}}\right)^2 \ln{10}\right)},10^{\frac{-A_m}{10}}\right\}.
\end{equation}
\par 
The array responses of the individual radiation
elements with each entry given as
\begin{equation}
\label{Eq:V}  [\mathbf{V}]_j=\exp{(i\mathbf{k}\mathbf{\cdot}\mathbf{x}_j)}
\end{equation}
where $\mathbf{\cdot}$ is the scalar dot product,  $\mathbf{x}_j$ is the location vector of the $j^{th}$ element, and $\mathbf{k}$ is the transmitted wave vector, where $\mathbf{k}=k\hat{\mathbf{v}}$, with $\hat{\mathbf{v}}$ being the unit wave vector and $k=\frac{2\pi}{\lambda}$. 
\subsubsection{Antenna Element on the curved surface}
The antenna elements mounted on the curved (lateral) surface of the cylinder are arranged along 
$N_v$ parallel circular rings stacked along the cylinder's vertical axis. 
Each ring contains $N_h$ elements uniformly distributed along the circumference. 
Let $s = 1, \ldots, N_h$ denote the index of an antenna column, where each column corresponds 
to the set of vertically aligned elements located at the same azimuthal angle. 
The boresight azimuth direction of the elements in the $s$-th column is defined as
$
{\psi}_s = \frac{2\pi (s-1)}{N_h}.$
Let $z = 1, \ldots, N_v$ denote the index of the circular ring along the vertical axis. 
Then, the horizontal radiation pattern of the element located in the $s$-th column and 
$z$-th ring is expressed as
\begin{equation}
    g_{E,H}^{(z,s)}(\phi) = g_{E,H}(\phi - {\psi}_s),
\end{equation}
Accordingly, its  overall radiation pattern with respect to the global coordinate system is given by:
\begin{equation}
F^{(z,s)}(\phi,\theta):= g_{E}^{(z,s)}(\phi, \theta)=g_{E,H}(\phi-{\psi}_s)g_{E,V}(\theta)
\end{equation}
We now consider the element with vertical index $z$ in the $s^{\text{th}}$ column of the cylindrical array. The corresponding array response entry is given by $$[\mathbf{V}]_{z,s}=\exp\!\left\{ \imath k\!\left( R\cos(\phi-\psi_s)\sin\theta +(z-1)d_v\cos\theta \right)\right\},$$ 
where $\psi_s=\frac{2\pi(s-1)}{N_h}$, $s=1,\ldots,N_h$. To derive this expression, we express the unit wave vector $\hat{\mathbf{v}}$ as: $\hat{\mathbf{v}}=(\cos\phi\sin\theta,\;\sin\phi\sin\theta,\;\cos\theta)$, and the position vector of the antenna element indexed by $(z,s)$, as: $\mathbf{x}_{z,s}=(R\cos\psi_s,\;R\sin\psi_s,\;(z-1)d_v)$.% The phase term follows from the plane-wave model $\exp\big(ik\,\hat{\mathbf{v}}^{T}\mathbf{x}_{z,s}\big)$, which simplifies to the expression above.

%Specifically, we know that the coordinate of $\hat{\mathbf{v}}=(\cos\phi\sin\theta, \sin\phi\sin\theta, \cos\theta)$  and the coordinate of $(z,s)^{th}$ element is $\mathbf{x}_{z,s}=(R\cos\hat{\psi}_s, R\sin\hat{\psi}_s, (z-1)d_v)$. 
\subsubsection{Antenna elements mounted on the bottom face of the cylinder}
\begin{figure}[h]
\centering
\includegraphics[width=3in]{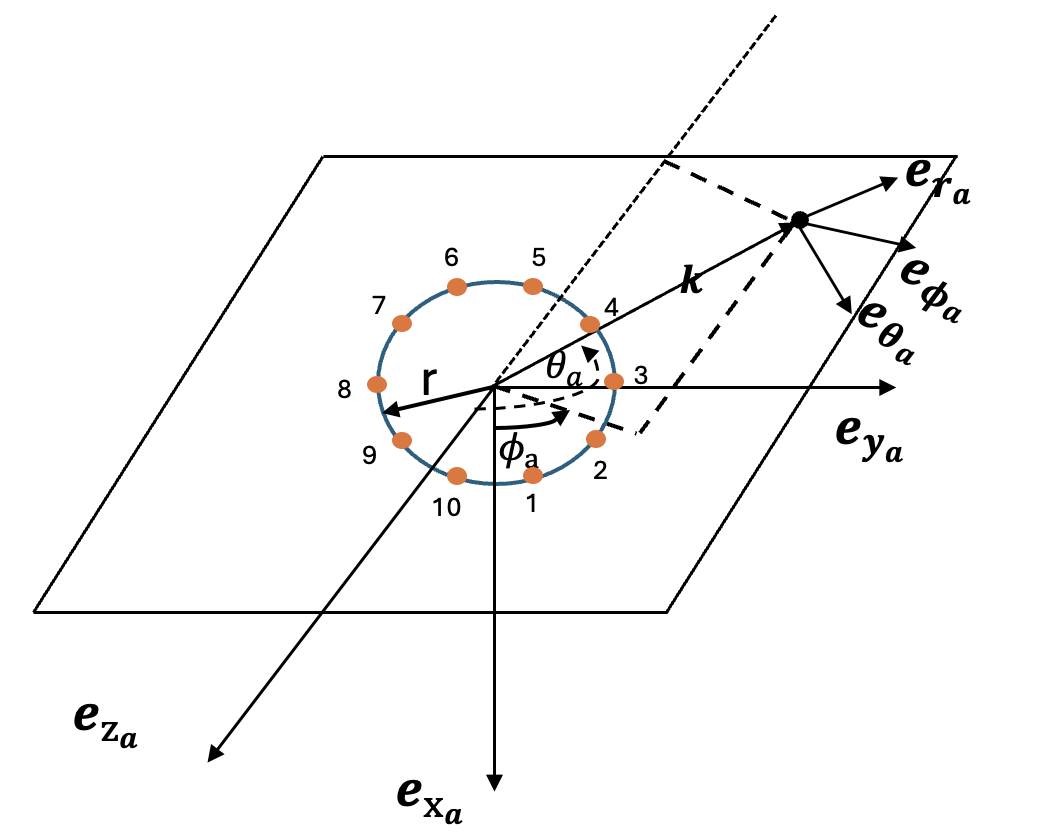}
\caption{Antenna configuration of HAPS Bottom}
\label{Fig: AN_Tx_circular}
\end{figure}

As for the elements mounted on the bottom face, we assume that they are uniformly distributed 
along a circle of radius $r$, with their boresight oriented toward the direction $-{\bf e}_z$. 
To achieve this configuration, each antenna element is rotated about the ${\bf e}_y$ axis, 
such that its local axis is aligned with ${\bf e}_{z_a}:={\bf e}_x$, as shown in Fig.\ref{Fig: AN_Tx_circular}. Since each antenna has been rotated, we endow it with a local spherical coordinate system 
in which the azimuth and elevation angles are defined with respect to its own orientation. 
If  $({\bf e}_{x_a}, {\bf e}_{y_a}, {\bf e}_{z_a})$ denote the local orthonormal basis 
associated with the antenna element, then these are related to the global coordinate system as :
$
    {\bf e}_{z_a} = {\bf e}_x, \quad
    {\bf e}_{x_a} = -{\bf e}_z, \quad \text{and} \ 
    {\bf e}_{y_a} = {\bf e}_y.$

Let $F^{(b)}(\phi,\theta)$ denote the radiation pattern of an antenna element mounted on the bottom face, evaluated in the direction specified by the azimuth angle $\phi$ and elevation angle $\theta$ with respect to the global coordinate system.
Then,
\begin{equation}
    F^{(b)}(\phi,\theta) = g_E(\phi_a, \theta_a),
\end{equation}
where $(\phi_a, \theta_a)$ are the azimuth and elevation angles expressed 
with respect to the local coordinate system of the antenna element and are given by:
\begin{align}
\theta_a&=\arccos({\bf e}_r.{\bf e}_{z})=\operatorname{arcos}(\cos \phi \sin \theta)\\
\phi_a =&\arg \left(\vec{e}_r \cdot \vec{e}_{x_a}+\jmath \vec{e}_r \cdot \vec{e}_{y_a}\right)= \tan^{-1} \left( \frac{ \mathbf{e}_r \cdot \mathbf{e}_{y_a} }{ \mathbf{e}_r \cdot \mathbf{e}_{x_a} } \right)  \nonumber \\
=& \tan^{-1} \left(\frac{\sin\phi\sin\theta}{-\cos \theta}\right).
\end{align}

For the antenna elements located at the bottom of the cylindrical array, i.e., the UCA configuration of radius $r$ depicted in Fig.~\ref{Fig: AN_Tx}, we denote by $\tilde{\mathbf{v}}\in\mathbb{C}^{N_b\times 1}$ the corresponding array response vector. This vector is expressed as:
\begin{equation}
\label{Eq:V_B}
[\tilde{\mathbf{v}}]_{b}= \exp{\left(\imath kr\cos(\phi-\hat{\psi}_b)\sin \theta \right)},
\end{equation}
where $\hat{\psi}_b=\frac{2\pi(b-1)}{N_b}, b=1,\cdots, N_b$. 

\section{Spatial correlation function}
\label{section: SCF}
\subsection{Radiation pattern correlation}
In this section, we derive the radiation pattern correlation for the non-line-of-sight (NLOS) components of the Rician channel model.
% Before deriving the radiation pattern correlation, it is essential to first describe the underlying system model to which it applies. We consider a multi-user communication system in which a HAPS is equipped with a cylindrical antenna system, as illustrated in Fig.~\ref{Fig: AN_Tx}. The HAPS serves $K$ users, each equipped with a single antenna and located at azimuth–zenith coordinates $(\phi_{k}^{\mathrm{LOS}},\,\theta_{k}^{\mathrm{LOS}})$ relative to the global coordinate frame of the HAPS. 
 We consider a HAPS equipped with a cylindrical antenna system, as illustrated in Fig.~\ref{Fig: AN_Tx}, to serve one user.
The propagation channel between the HAPS and user  is modeled as a \emph{Rician fading channel}, decomposed into line-of-sight (LOS) and NLOS components:
\begin{equation}
{\bf h}
= \sqrt{\beta}\!\left(
\sqrt{\tfrac{\kappa}{\kappa+1}}\,{\bf h}^{\mathrm{LOS}}
+ 
\sqrt{\tfrac{1}{\kappa+1}}\,{\bf h}^{\mathrm{NLOS}}
\right),
\label{eq:rician_model}
\end{equation}
with $ \beta_{\mathrm{dB}} =P_t + G_{\mathrm{UE}} - PL_{\mathrm{dB}} + G_{E,\mathrm{max}},$
where $P_t$ denotes the user's transmit power, $G_{\mathrm{UE}}$  the user's antenna gain, $PL_{\mathrm{dB}}$  the free-space path loss in dB, and $G_{E,\mathrm{max}}$  the maximum element gain of the transmit antenna array. In addition, $\kappa$ denotes the Rician factor representing the relative strength of the LOS component.
\par 
The LOS channel entry corresponding to antenna 
$p$ for user is given by
\begin{equation}
\begin{aligned}
[{\bf h}^{\mathrm{LOS}}]_p
=&
\exp\!\Big(-\jmath \tfrac{2\pi d_{3\mathrm{D}}}{\lambda}\Big)
\sqrt{F_p(\phi^{\mathrm{LOS}},\theta^{\mathrm{LOS}})}\,\\
&\times[{\bf a}_{\mathrm{tx}}(\phi^{\mathrm{LOS}},\theta^{\mathrm{LOS}})]_p,
\label{eq:los_component}
\end{aligned}
\end{equation}
where $d_{3\mathrm{D}}$ denotes the 3D distance between the HAPS and user, $\lambda$ is the carrier wavelength,  and  the radiation pattern and the array-response entry of antenna 
$p$ are specified as:
\begin{equation}
F_p(\phi^n,\theta^n)=
\begin{cases}
F^{(z,s)}(\phi^n,\theta^n), & \text{if } p\in \mathcal{S}_{\mathrm{cyl}},\\[2pt]
F^{(b)}(\phi^{n},\theta^n),   & \text{if } p\in \mathcal{S}_{\mathrm{bot}},
\end{cases}
\label{eq:pattern}
\end{equation}

\begin{equation}
[{\bf a}_{\mathrm{tx}}(\phi^{n},\theta^{n})]_p=
\begin{cases}
{\bf V}_{z,s},      & \text{if } p\in \mathcal{S}_{\mathrm{cyl}},\\[2pt]
[\tilde{\bf v}]_b,  & \text{if } p\in \mathcal{S}_{\mathrm{bot}}.
\end{cases}
\label{eq:array_response}
\end{equation}
Here, $(z,s)$ denote the vertical and circumferential indices of antenna element $p$ when it is located on the curved surface, whereas $b$ denotes the index of an element located on the bottom surface.
The NLOS component with respect to antenna $p$ aggregates $N$ scattered paths as
\begin{equation}
[{\bf h}^{\mathrm{NLOS}}]_p
=
\sum_{n=1}^{N}
\alpha_n
\sqrt{F_p(\phi^n,\theta^n)}\\,
[{\bf a}_{\mathrm{tx}}(\phi^{n},\theta^{n})]_p,
\label{eq:nlos_component}
\end{equation}
where $\alpha_n$ denotes the complex amplitude of the $n$-th path, respectively, and $(\phi^{n},\theta^{n})$ are its corresponding azimuth and zenith angles of depature. Additionally, the complex amplitudes are assumed to be i.i.d zero mean, $\frac{1}{N}$ variance Gaussian random variables.
\par
 We are interested in computing the spatial correlation function between the NLOS components of antenna elements situated either on the bottom surface or on the curved surface. Owing to the differences in antenna boresight orientation and spatial separation, the NLOS components of elements on the bottom surface can be safely assumed to be uncorrelated with those on the curved surface.
\par
Let $p$ and $q$ denote the indices of two antenna elements located either on the curved surface or on the bottom surface. 
% We assume that, across the $N$ propagation paths,
% the random phases $\{\Phi_l^{n}\}_{n=1}^{N}$ are mutually independent and independent of the azimuth and zenith angles associated with the paths. 
We assume that the complex amplitudes  associated with each path are independent of the angular parameters. Hence, 
the spatial correlation between the the correlation between $[{\bf h}^{\mathrm{NLOS}}]_p$ and $[{\bf h}^{\mathrm{NLOS}}]_q$
is given by
\begin{equation}
\begin{aligned}
&\mathbb{E}\!\left[[{\bf h}^{\mathrm{NLOS}}]_p
[{\bf h}^{\mathrm{NLOS}}]_q^{\!*}\right]\nonumber\\
&= \mathbb{E}\!\big[
\sqrt{F_p(\phi, \theta)}\sqrt{F_q(\phi, \theta)}
[{\bf a}_{\mathrm{tx}}(\phi, \theta)]_p
[{\bf a}_{\mathrm{tx}}(\phi, \theta)]_q^\star
\big].
\end{aligned}
\label{eq:SCF}
\end{equation}
It appears from the above equation that a fundamental building block in the characterization 
of the spatial correlation between the NLOS components is the radiation pattern correlation. 
The radiation pattern correlation quantifies the statistical dependence between the NLOS components observed at 
different antenna elements as a function of their relative geometry, their gains, and  the angular distribution.
% and is given as:
% $$
% \mathbb{E}\Big[\sqrt{F_p(\phi, \theta)}\sqrt{F_q(\phi, \theta)}[{\bf a}_{\mathrm{tx}}(\phi, \theta)]_p
% [{\bf a}_{\mathrm{tx}}(\phi, \theta)]_q^\star\Big]
% $$
\par
Accordingly, the radiation pattern correlation between two antenna elements $p$ and $q$, both located on the curved surface with circumferential and vertical indices $(z,s)$ and $(z',s')$, respectively, is defined as
\begin{equation}
\begin{aligned}
&\rho\big((z,s),(z',s')\big)\\
&= \mathbb{E}\!\left[
\sqrt{F^{(z,s)}(\phi, \theta)}\, 
\sqrt{F^{(z',s')}(\phi, \theta)}\,
[{\bf V}]_{z,s}\,
[{\bf V}]_{z',s'}^{\!*}
\right],
\end{aligned}
\label{eq:SCF_cyl}
\end{equation}
where the expectation is taken with respect to the joint distribution of the azimuth and zenith angles $(\phi,\theta)$. Similarly, the radiation pattern correlation between two antenna elements $p$ and $q$ located on the bottom surface is defined as:
\begin{equation}
\rho_b({b,b'})
= \mathbb{E}\!\left[
\sqrt{F^{(b)}(\phi, \theta)}\sqrt{F^{(b)}(\phi_l, \theta_l)}\,
[\tilde{\bf v}]_b\,
[\tilde{\bf v}]_{b'}^{\star}
\right],
\label{eq:SCF_bot}
\end{equation}
In the sequel, we will sequentially derive closed-form expressions for these radiation pattern correlation quantities.
\subsubsection{Radiation pattern correlation for  elements on the curved surface}
Let two antenna elements on the curved surface with circumferential and vertical indices $(z,s)$ and $(z',s')$. Then, the radiation pattern correlation between these elements is written as:
\begin{align}
&\rho((z,s),(z',s'))\nonumber \\
&=\mathbb{E}\big[g_{E,H}^{s,s'}(\phi)g_{E,V}(\theta)\Big(\exp(\imath k(R\sin\theta (\cos(\phi-{\psi}_s)-\cos(\phi-{\psi}_{s'})) \nonumber \\
&\times d_v(z-z')\cos\theta)\Big)\big]
\end{align}
where 
$$
g_{E,H}^{s,s'}=\sqrt{g_{E,H}(\phi-{\psi}_s)}\sqrt{g_{E,H}(\phi-{\psi}_s^{'})}
$$
Define:
\begin{align}
\rho_1(s,s')&=\exp\Big(\imath kR\sin\theta\big(\cos(\phi-{\psi}_s)-\cos(\phi-{\psi}_{s'})\big)\Big)\\
\rho_2(z,z')&=\exp(\imath k d_v \cos\theta (z-z'))
\end{align}
Then,
$$
\rho((z,s),(z',s'))=\mathbb{E}\big[g_{E,H}^{s,s'}(\phi)g_{E,V}(\theta)\rho_1(s,s')\rho_2(z,z')\big].
$$
To continue, we simplify the expression of $\rho_1(s,s')$ and $\rho_2(z,z')$. For that, we let $Z_1^{s,s'}=\cos({\psi}_s)-\cos({\psi}_{s'})$, $Z_2^{s,s'}=\sin({\psi}_s)-\sin({\psi}_{s'})$, $Z_3^{s,s'}=R\sqrt{(Z_1^{s,s'})^2+(Z_{2}^{s,s'})^2}$, $Z_4^{z,z'}=d_v(z-z')$ and $a_{s,s'}^{z,z'}=\sqrt{(Z_3^{s,s'})^2+(Z_4^{z,z'})^2}$.
Then, using these notations, we obtain:
\begin{align}
\rho_1(s,s')&=\exp(\imath kR\sin\theta(Z_1^{s,s'}\cos\phi+Z_2^{s,s'}\sin\phi))\nonumber\\
&=\exp(\imath kR \sin \theta \sqrt{(Z_1^{s,s'})^2+(Z_2^{s,s'})^2}\cos(\phi-\zeta_{s,s'}))\nonumber\\
&=
\exp(\imath ka_{s,s'}^{z,z^{'}}\sin \theta \sin\beta_{s,s'}^{z,z'}\cos(\phi-\zeta_{s,s^{\prime}}))
\end{align}
where $\beta_{s,s'}^{z,z'}=\tan^{-1}(Z_3^{s,s'}/Z_4^{z,z'})$ and $\zeta_{s,s'}=\tan^{-1}(Z_2^{s,s'}/Z_1^{s,s'})$.
Similarly, we have
$$
\rho_2(z,z')=\exp(ika_{s,s'}^{z,z'}\cos\beta_{s,s'}^{z,z'} \cos\theta)
$$
With this we obtain:
\begin{align}
&\rho((z,s),(z',s'))=\mathbb{E}\big[g_{E,H}^{s,s'}(\phi)g_{E,V}(\theta)\nonumber \\
&\times\exp(\imath ka_{s,s'}^{z,z'}(\sin\beta_{s,s'}^{z,z'}\sin \theta\cos(\phi-\zeta_{s,s'})+\cos\beta_{s,s'}^{z,z'}\cos\theta))\big]
\end{align}
We can interpret $\sin\beta_{s,s'}^{z,z'}\sin \theta\cos(\phi-\zeta_{s,s'})+\cos\beta_{s,s'}^{z,z'}\cos\theta$ as the scalar product between the product of the unit vector $\hat{\bf v}$ with azimuth $\phi$ and zenith $\theta$ and the unit vector $\hat{\bf x}_{s,s'}^{z,z'}$ with azimuth $\zeta_{s,s'}$ and zenith $\beta_{s,s'}^{z,z'}$. This allows us to simplify the pattern correlation function as:
\begin{align}
\rho((z,s),(z',s'))&=\mathbb{E}\big[g_{E,H}^{s,s'}(\phi)g_{E,V}(\theta)\nonumber \\
&\times\exp(\imath ka_{s,s'}^{z,z'}(({\bf x}_{s,s'}^{z,z'})^{T}\hat{\bf v})\big]
\end{align}
and then to use the spherical harmonics expansion and the addition theorem in Lemma \ref{lemma:lemma1} in Appendix \ref{app:A} to find:
\begin{equation}
\begin{aligned}
&\rho((z,s), (z', s')) \nonumber \\
&=\mathbb{E}\Bigg[g_ {E,H}^{s,s'} ( \phi , \theta )g_{E,V}(\theta)\sum_{n=0}^{\infty} i^n(2 n+1) j_n\big(\frac{2 \pi}{\lambda} a_{s,s'}^{z,z'}\big)\\
& \times\Big(P_n(\cos \theta) P_n(\cos \beta_{s,s'}^{z,z'})+2 \sum_{m=1}^n \frac{(n-m)!}{(n+m)!} P_n^m(\cos \theta) \\
& \times P_n^m(\cos \beta_{s,s'}^{z,z'}) \cos (m(\phi-\zeta_{z,z'}))\Big)\Bigg]
\end{aligned}
\end{equation}
where $P_n(\cdot)$ and $P_n^m(\cdot)$ denote the Legendre polynomials of order $n$ and their associated Legendre functions, respectively. Defining $
\bar{P}_n^m(x) = \sqrt{\left(n + \tfrac{1}{2}\right) \frac{(n - m)!}{(n + m)!}} \, P_n^m(x)
$, 
we rewrite $\rho((z, s), (z', s'))$ in the form given by equation~\eqref{eq:rho3}, which appears at the top of the next page.
The expression in \eqref{eq:rho3}  does not lend itself to a direct derivation of a closed-form expression for the radiation pattern correlation, as the azimuth and zenith angles appear within the arguments of the Legendre polynomials.  To overcome this difficulty, we employ the trigonometric expansion of the Legendre polynomials  and associated Legendre polynomials presented in Lemma~\ref{lemma:lemma2} in  Appendix \ref{app:A}. This expansion enables us to express $\rho((z, s), (z', s'))$ in terms of the Fourier spectrum coefficients of the power azimuth spectrum (PAS) and the power elevation spectrum (PES), defined as follows:
\begin{equation}
\operatorname{PAS}_E^{s,s'}(\phi) = g_{E,H}^{s,s'}(\phi)\, p_\phi(\phi),
\end{equation}
\begin{equation}
\operatorname{PES}_E(\theta) = g_{E,V}(\theta)\, p_\theta(\theta),
\end{equation}
where the angular power density functions $p_\phi(\phi)$ and $p_\theta(\theta)$ are related to the probability density functions of the azimuth and elevation angles, $f_\phi(\phi)$ and $f_\theta(\theta)$, through
\[
p_\phi(\phi) = f_\phi(\phi), \qquad 
p_\theta(\theta) = \frac{f_\theta(\theta)}{\sin(\theta)}.
\]
Define the Fourier series (FS) coefficients $a_\phi^{s,s'}(m)$, $b_\phi^{s,s'}(m)$, $a_\theta(k)$, and $b_\theta(k)$ of the respective power spectra  as:
\begin{equation}
a_\phi^{s,s'}(m) = \frac{1}{\pi} \int_{-\pi}^{\pi} \operatorname{PAS}_E^{s,s'}(\phi)\, \cos(m\phi)\, d\phi,
\end{equation}
\begin{equation}
b_\phi^{s,s'}(m) = \frac{1}{\pi} \int_{-\pi}^{\pi} \operatorname{PAS}_E^{s,s'}(\phi)\, \sin(m\phi)\, d\phi,
\end{equation}
\begin{equation}
a_\theta(k) = \frac{1}{\pi} \int_{0}^{\pi} \operatorname{PES}_E(\theta)\, \cos(k\theta)\, d\theta,
\end{equation}
\begin{equation}
b_\theta(k) = \frac{1}{\pi} \int_{0}^{\pi} \operatorname{PES}_E(\theta)\, \sin(k\theta)\, d\theta.
\end{equation}
We thus obtain the final analytical expression for the radiation pattern correlation in \eqref{Eq:rho_final}. 
The main advantage of this approach is that it provides a systematic framework for evaluating the radiation pattern correlation for arbitrary azimuth and zenith angle distributions, and for any antenna gain patterns, solely as a function of the Fourier coefficients of the corresponding power spectra in zenith and azimuth \cite{kammoun2015generalized}.

\begin{figure*}[h]
 \begin{equation}
 \label{eq:rho3}
 \begin{aligned}
 \rho((z,s), (z', s'))= & \mathbb{E}[g_{E,H}^{s,s'}(\phi) g_{E,V}(\theta)]j_0(\frac{2 \pi}{\lambda} a_{s,s'}^{z,z'})+ \sum_{n=1}^{\infty} (-1)^{n}(4n+1) j_{2n}(\frac{2 \pi}{\lambda} a_{s,s'}^{z,z'})P_{2n}(\cos \beta_{s,s'}^{z,z'})\mathbb{E}[P_{2n}(\cos \theta)g_{E, V}(\theta)]
 \mathbb{E}[g_{E, H}^{s,s'}(\phi)] \\
 & -\sum_{n=1}^{\infty} \imath(-1)^{n}(4n-1) j_{2n-1}\big(\frac{2 \pi}{\lambda} a_{s,s'}^{z,z'}\big) P_{2n-1}(\cos \beta_{s,s'}^{z,z'})  \mathbb{E}[P_{2n-1}(\cos \theta)g_ {E,V}(\theta)]\mathbb{E}[g_ {E,H}^{s,s'}(\phi)]\\
 &+ \sum_{n=1}^{\infty} 4(-1)^{n} j_{2n}\big(\frac{2 \pi}{\lambda} a_{s,s'}^{z,z'} \big)\Big(\sum_{m=1}^{n}\bar{P}_{2n}^{2m}(\cos \beta_{s,s'}^{z,z'}) \mathbb{E}\big[\bar{P}_{2n}^{2m}(\cos \theta)g_{E,V}(\theta)\big]  \mathbb{E}\big[\cos (2m(\phi-\zeta_{s,s'}))g_{E,H}^{s,s'}(\phi)\big] \\
 &+ \sum_{m=1}^{n}\bar{P}_{2n}^{2m-1}(\cos \beta_{s,s'}^{z,z'}) \mathbb{E}[\bar{P}_{2n}^{2m-1}(\cos \theta)g_{E,V}(\theta)]  \mathbb{E}[\cos [(2m-1)(\phi-\zeta_{s,s'})]g_{E,H}(\phi)] \Big)\\
 &-\sum_{n=1}^{\infty} \imath(-1)^{n}4 j_{2n-1}\big(\frac{2 \pi}{\lambda} a_{s,s'}^{z,z'}\big)  \Big(\sum_{m=1}^{n}\bar{P}_{2n-1}^{2m}(\cos \beta_{s,s'}^{z,z'})\mathbb{E}[ \bar{P}_{2n-1}^{2m}(\cos \theta)g_{E,V}(\theta)] \mathbb{E}\big[\cos (2m(\phi-\zeta_{s,s'}))g_{E,H}(\phi)\big] \\
 & +\sum_{m=1}^{n}\bar{P}_{2n-1}^{2m-1}(\cos \beta_{s,s'}^{z,z'})\mathbb{E}[\bar{P}_{2n-1}^{2m-1}(\cos \theta)g_{E,V}(\theta)]\mathbb{E}\big[\cos ((2m-1)(\phi-\zeta_{s,s'}))g_{E,H}(\phi)\big]\Big)
 \end{aligned}
 \end{equation}
 \end{figure*}
% \par
% To further simplify $\rho(s, s^{\prime}, z, z^{\prime})$, we explore the FS coefficients of the power azimuth spectrum (PAS) and the power elevation spectrum (PES), which defined as \cite{kammoun2016spatial}:
% \begin{equation}
% \operatorname{PAS}_E(\phi)=g_{E, H}(\phi) p_\phi(\phi)
% \end{equation}
% \begin{equation}
%     \operatorname{PES}_E(\theta)=g_{E, V}(\theta) p_\theta(\theta)
% \end{equation}
% where the angular power density functions $p_\phi(\phi)$ and $p_\theta(\theta)$ equal $f_{\phi}(\phi)$ and $\frac{f_\theta(\theta)}{\sin (\theta)}$ respectively, with $f_{\phi}(\phi)$ and $f_\theta(\theta)$ being the probability density functions of the azimuth and elevation angles. The FS coefficients, $a_\phi(m), b_\phi(m), a_\theta(k)$ and $b_{\theta}(k)$, for the power spectra are defined as,
% \begin{equation}
% a_\phi(m)=\frac{1}{\pi} \int_{-\pi}^\pi \operatorname{PAS}_E(\phi) \cos (m\phi) d \phi
% \end{equation}
% \begin{equation}
% b_\phi(m)=\frac{1}{\pi} \int_{-\pi}^\pi \operatorname{PAS}_E(\phi) \sin (m\phi) d \phi
% \end{equation}
% \begin{equation}
% a_\theta(k)=\frac{1}{\pi} \int_0^\pi \operatorname{PES}_E(\theta) \cos (k \theta) d \theta
% \end{equation}
% \begin{equation}
% b_\theta(k)=\frac{1}{\pi} \int_0^\pi \operatorname{PES}_E(\theta) \sin (k \theta) d \theta
% \end{equation}
% Therefore, combining the trigonometric expansion of Legendre polynomials
% summarized in the following Lemma \ref{lemma: lemma2}, the final analytical expression for the SCF is presented in equation (\ref{Eq:rho_final}), as shown on the next page.
\begin{figure*}[h]
\begin{small}
\begin{equation}
\label{Eq:rho_final}
\begin{aligned}
&\rho((z, s), (z', s'))=  \pi^2a_{\phi}^{s,s'}(0)b_{\theta}(1) j_0(\frac{2 \pi}{\lambda} a_{s,s'}^{z,z'})\\
&+ \sum_{n=1}^{\infty} (-1)^{n}(4n+1) j_{2n}\big(\frac{2 \pi}{\lambda} a_{s,s'}^{z,z'}\big)P_{2n}(\cos \beta_{s,s'}^{z,z'})\pi^2 a_{\phi}^{s,s'}(0)
\sum_{k=-n}^n p_{n-k} p_{n+k} 
\frac{1}{2}\left(b_{\theta}(2k+1)-b_{\theta}(2k-1)\right) \\
& -\sum_{n=1}^{\infty} \imath(-1)^{n}(4n-1) j_{2n-1}\big(\frac{2 \pi}{\lambda} a_{s,s'}^{z,z'}\big) P_{2n-1}(\cos \beta_{s,s'}^{z,z'}) \pi^2 a_{\phi}^{s,s'}(0) \sum_{k=1}^n p_{n-k} p_{n+k-1} \left(b_{\theta}(2k)-b_{\theta}(2k-2)\right)\\
&+ \sum_{n=1}^{\infty} 4(-1)^{n} j_{2n}\big(\frac{2 \pi}{\lambda} a_{s,s'}^{z,z'}\big)\left(\sum_{m=1}^{n}\bar{P}_{2n}^{2m}(\cos \beta_{s,s'}^{z,z'}) \frac{\pi^2 }{2}(\cos (2 m \zeta_{s,s^{\prime}}) a_\phi^{s,s'}(2 m)+\sin (2 m \zeta_{s,s'}) b_\phi^{s,s'}(2 m)) \sum_{k=0}^n  c_{2 n, 2 k}^{2 m} \left[b_{\theta}(2k+1)-b_{\theta}(2k-1)\right] \right.\\
&\left.+ \sum_{m=1}^{n}\bar{P}_{2n}^{2m-1}(\cos \beta_{s,s'}^{z,z'}) \frac{\pi^2}{2}   (\cos ((2 m-1) \zeta_{s,s^{\prime}}) a_\phi^{s,s'}(2 m-1) +\sin ((2 m-1) \zeta_{s,s'}) b_\phi^{s,s'}(2 m-1)) \sum_{k=1}^n  d_{2 n, 2 k}^{2 m-1} \left(a_{\theta}(2k-1)-a_{\theta}(2k+1)\right)  \right)\\
&-\sum_{n=1}^{\infty} \imath(-1)^{n}4 j_{2n-1}\big(\frac{2 \pi}{\lambda} a_{s,s'}^{z,z'}\big)  \left(\sum_{m=1}^{n}\bar{P}_{2n-1}^{2m}(\cos \beta_{s,s'}^{z,z'}) \frac{\pi^2}{2} (\cos (2 m \zeta_{s,s'}) a_\phi^{s,s'}(2 m)+\sin (2 m \zeta_{s,s'}) b_\phi^{s,s'}(2 m)) \sum_{k=1}^n  c_{2 n-1,2 k-1}^{2 m} \left(b_{\theta}(2k)-b_{\theta}(2k-2)\right) \right.\\
&\left. +\sum_{m=1}^{n}\bar{P}_{2n-1}^{2m-1}(\cos \beta_{s,s'}^{z,z'}) \frac{\pi^2}{2}  (\cos ((2 m-1) \zeta_{s,s'}) a_\phi^{s,s'}(2 m-1) +\sin ((2 m-1) \zeta_{s,s'}) b_\phi^{s,s'}(2 m-1))\sum_{k=1}^n d_{2 n-1,2 k-1}^{2 m-1}  \left(a_{\theta}(2k-2)-a_{\theta}(2k)\right)\right)
\end{aligned}
\end{equation}
\end{small}
\end{figure*}
\par

\subsubsection{Radiation pattern correlation for elements on the bottom face}
Let two antenna elements on the bottom face with indexes $b$ and $b'$. Then, the radiation pattern correlation between these elements  is given by:
\begin{equation*}
\begin{aligned}
\rho_b(b,b')=&\mathbb{E}\Big[g_E(\phi_a,\theta_a)\exp(\imath kr\sin\theta (\cos(\phi-\hat{\psi}_b)\\
&-\cos(\phi-\hat{\psi}_{b'})))\Big]
\end{aligned}
\end{equation*}
Let $\hat{Z}_1^{b,b'}=\cos(\hat{\psi}_b)-\cos(\hat{\psi}_{b'})$ and $\hat{Z}_2^{b,b'}=\sin \hat{\psi}_b-\sin \hat{\psi}_{b'}$
Then, $\rho(b,b')$ simplifies as:
\begin{align}
\rho_b(b,b')&=\mathbb{E}\big[g_{E}(\phi_a,\theta_a)\exp(\imath k r \sin(\theta)\nonumber\\
&\times(\hat{Z}_{1}^{b,b'}\cos\phi+\hat{Z}_2^{b,b'}\sin\phi))\big] \nonumber\\
&=\mathbb{E}\big[g_E(\phi_a,\theta_a)\exp(\imath k r \hat{c}_{b,b'} \sin(\theta)\cos(\phi-\hat{\zeta}_{b,b'})\big]
\end{align}
where $\hat{\zeta}_{b,b'}=\tan^{-1}(\hat{Z}_2^{b,b'}/\hat{Z}_1^{b,b'})$ and $\hat{c}_{b,b'}=\sqrt{(\hat{Z}_1^{b,b'})^2+(\hat{Z}_2^{b,b'})^2}$. 
Similarly as in the case of elements on the curved surface, $\sin\theta\cos(\phi-\hat\zeta_{b,b'})$ can be interpreted as the scalar product between the unit vector $\hat{\bf v}$ with azimuth $\phi$ and elevation $\theta$ and the vector $\hat{\bf x}_{b,b'}$ with azimuth $\hat{\zeta}_{b,b'}$ and zenith $\frac{\pi}{2}$. 
By applying Lemma \ref{lemma:lemma1} in Appendix \ref{app:A}, we thus obtain:
\begin{equation}
\begin{aligned}
\rho_b(b,b') & =\mathbb{E}\Big[g_E ( \phi_a , \theta_a )\sum_{n=0}^{\infty} \imath^n(2 n+1) j_n\big(\frac{2 \pi}{\lambda} r\hat{c}_{b,b'}\big) \\
& \Big(P_n(\cos \theta) P_n(0)+2 \sum_{m=1}^n \frac{(n-m)!}{(n+m)!} P_n^m(\cos \theta) \\
& \times P_n^m(0) \cos \big(m(\phi-\hat{\zeta}_{b,b'})\big)\Big)\Big]
\end{aligned}
\end{equation}
By following the same methodology as for the elements located on the curved surface, 
we expand the pattern correlation using the trigonometric representation of the Legendre polynomial in \eqref{eq:rho_b2}. 
The key difference, however, is that the azimuth and zenith angles are no longer separable in this case. 
As a result, the final expression depends on the two-dimensional power spectrum
${\rm PS}(\phi,\theta)=g_E(\phi_a,\theta_a)\,f_\phi(\phi)\,f_\theta(\theta)$ 
through its associated two-dimensional Fourier coefficients defined as:
\begin{equation}
\tilde{a}(m,n)\!=\!\frac{1}{\pi^2} \!\!\!\int_{0}^{\pi}\!\!\!\int_{-\pi}^\pi \!\! \! g_E(\phi_a,\theta_a) f_{\phi}(\phi)f_{\theta}(\theta) \cos (m\theta)\cos(n\phi)d \phi d \theta
\end{equation}
\begin{equation}
\tilde{b}(m,n)=\!\frac{1}{\pi^2}\!\!\! \int_{0}^{\pi} \!\!\!\int_{-\pi}^\pi\!\!\! g_E( \phi_a,\theta_a) f_{\phi}(\phi)f_{\theta}(\theta) \sin (m \theta) \sin(n\phi)d \phi d\theta
\end{equation}
\begin{equation}
\tilde{c}(m,n)\!\!=\!\!\frac{1}{\pi^2} \!\!\!\int_0^\pi\!\!\!\int_{-\pi}^\pi \!\!\!g_E(\phi_a,\theta_a)  f_{\phi}(\phi)f_{\theta}(\theta) \cos (m \theta)\sin(n\phi)d\phi d \theta
\end{equation} 
\begin{equation}
\tilde{d}(m,n)\!\!=\!\!\frac{1}{\pi^2}\!\!\! \int_0^\pi\!\!\! \int_{-\pi}^\pi \!\!\! g_E(\phi_a,\theta_a) f_{\phi}(\phi)f_{\theta}(\theta)\sin (m\theta)\cos(n\phi) d \phi  d \theta
\end{equation}
The final expression can be found in \eqref{eq: rhob3}. Similar to the curved-surface case, we note that the correlation expression for the bottom-surface elements is  generic, requiring only the Fourier coefficients of the corresponding power spectrum to handle arbitrary angle distributions and antenna gain patterns.

\begin{remark}
The proposed closed-form SCF expression is represented by an infinite series. We agree that, for practical implementation, the infinite summation must be truncated to a finite number of terms, denoted by \(N_0\). In the following, we show that the truncation error admits an exponentially decaying upper bound.
Let two antenna elements on the curved surface with circumferential and vertical indices $(z,s)$ and $(z',s')$. Then, the radiation pattern correlation between these elements is written as:
\begin{equation}
\begin{aligned}
&\rho((z,s), (z', s'))\!\!=\!\!\mathbb{E}\big[g_{E,H}^{s,s'}(\phi)g_{E,V}(\theta)\exp(\imath ka_{s,s'}^{z,z'}(({\mathbf x}_{s,s'}^{z,z'})^{T}\hat{\mathbf v})\big]\\
&\!\!=\!\!\sum_{n=0}^{\infty}i^n(2 n+1) j_n\big(\frac{2 \pi}{\lambda} a_{s,s'}^{z,z'}\big)\mathbb{E}\big[g_{E,H}^{s,s'}(\phi)g_{E,V}(\theta) P_n(({\mathbf x}_{s,s'}^{z,z'})^{T}\hat{\mathbf v})\big]
\end{aligned}
\end{equation}
After truncating the series at order \(N_0\), the resulting truncation error is
\begin{equation}
\begin{aligned}
\epsilon_{N_0}&=\sum_{n>N_0}i^n(2 n+1) j_n\big(\frac{2 \pi}{\lambda} a_{s,s'}^{z,z'}\big)\\
& \times \mathbb{E}\big[g_{E,H}^{s,s'}(\phi)g_{E,V}(\theta) P_n(({\mathbf x}_{s,s'}^{z,z'})^{T}\hat{\mathbf v})\big],\\
&\leq \sum_{n>N_0} (2 n+1) |j_n\big(\frac{2 \pi}{\lambda} a_{s,s'}^{z,z'}\big)\\
&\times  |\big|\mathbb{E}\big[g_{E,H}^{s,s'}(\phi)g_{E,V}(\theta) P_n(({\mathbf x}_{s,s'}^{z,z'})^{T}\hat{\mathbf v})\big]\big|,\\
& \leq \sum_{n>N_0} (2 n+1)| j_n\big(\frac{2 \pi}{\lambda} a_{s,s'}^{z,z'}\big)|,
\end{aligned}
\end{equation}
where the last inequality follows from $\sup_{\mathbf{x}\leq 1}|P_n(\mathbf{x})|\leq 1$ \cite{rivlin1981introduction} and $|\mathbb{E}\big[g_{E,H}^{s,s'}(\phi)g_{E,V}(\theta) \big]\big|\leq 1$. 
\par
It is straightforward to observe that the above series has the same form as Eq.~(10b) in \cite{abhayapala2003characterization}. Therefore, the truncation analysis in \cite{abhayapala2003characterization} can be directly applied. Specifically, from Eq.~(9) therein, the truncation error is bounded by
\begin{equation}
    \epsilon_{N_0}\leq \eta \exp(-\Delta),
\end{equation}
where $\Delta=N_0-\left\lceil eka_{s,s'}^{z,z'}/2\right\rceil$. To obtain a worst-case bound independent of the antenna pair indices, we further bound 
$
a_{s,s'}^{z,z'} \leq a_{\max} \triangleq d_v\sqrt{\frac{N_h^2}{\pi^2}+(N_v-1)^2}
$. For the considered array configuration, $d_v=d_h=0.5\lambda$, $N_h=16$ and $N_v=6$, we get
\begin{equation}
\left\lceil
\frac{e\pi}{2}
\sqrt{
\frac{N_h^2}{\pi^2}
+
(N_v-1)^2
}
\right\rceil
=31,
\end{equation}
which leads to
\begin{equation}
\Delta=N_0-31.
\end{equation}
Consequently,
\begin{equation}
\epsilon_{N_0}
\le
0.678481243\,e^{-(N_0-31)}.
\end{equation}
For the value used in this work, namely \(N_0=35\), we obtain
\begin{equation}
\epsilon_{35}
\le
0.678481243\,e^{-4}
\approx
1.24\times10^{-2}.
\end{equation}
This result confirms that the truncation error is sufficiently small for practical implementation and has a negligible impact on the numerical results reported in the paper. It should be noted that the same truncation analysis applies to the bottom-surface antenna elements. Since all bottom elements lie on the same horizontal plane (\(z=z'\)), the corresponding expressions are further simplified, while the truncation error remains exponentially decaying.
\end{remark}
 \begin{figure*}[h!]
 \begin{equation}
 \label{eq:rho_b2}
 \begin{aligned}
 \rho_b(b, b^{\prime})= & \mathbb{E}[g_E(\phi_a,\theta_a)]j_0(\frac{2 \pi}{\lambda} r\hat{c}_{b,b'})+ \sum_{n=1}^{\infty} (-1)^{n}(4n+1) j_{2n}\big(\frac{2 \pi}{\lambda} r\hat{c}_{b,b'}\big)P_{2n}(0)\mathbb{E}[P_{2n}(\cos \theta)g_E(\phi_a, \theta_a)] \\
 &+ \sum_{n=1}^{\infty} 4(-1)^{n} j_{2n}\big(\frac{2 \pi}{\lambda} r\hat{c}_{b,b'} \big)\left(\sum_{m=1}^{n}\bar{P}_{2n}^{2m}(0) \mathbb{E}\Big[\bar{P}_{2n}^{2m}(\cos \theta)\cos \big(2m(\phi-\hat{\zeta}_{b,b'})\big)g_E(\phi_a,\theta_a)\Big] \right)\\
 &-\sum_{n=1}^{\infty} 4\imath(-1)^{n} j_{2n-1}\big(\frac{2 \pi}{\lambda} r\hat{c}_{b,b'}\big)  \left(\sum_{m=1}^{n}\bar{P}_{2n-1}^{2m-1}(0)\mathbb{E}\Big[\bar{P}_{2n-1}^{2m-1}(\cos \theta)\cos \big((2m-1)(\phi-\hat{\zeta}_{b,b'})\big)g_E(\phi_a,\theta_a)\Big]\right)\\
 & =\mathbb{E}\big[g_E(\phi_a,\theta_a)\big]j_0(\frac{2 \pi}{\lambda} r\hat{c}_{b,b'})+ \sum_{n=1}^{\infty} (-1)^{n}(4n+1) j_{2n}\big(\frac{2 \pi}{\lambda} r\hat{c}_{b,b'}\big)P_{2n}(0)\mathbb{E}\big[P_{2n}(\cos \theta)g_E(\phi_a,\theta_a)\big] \\
 &+ \sum_{n=1}^{\infty} 4(-1)^{n} j_{2n}\big(\frac{2 \pi}{\lambda} r\hat{c}_{b,b'}\big)\left(\sum_{m=1}^{n}\bar{P}_{2n}^{2m}(0) \left(\cos(2m\hat{\zeta}_{b,b'})\mathbb{E}\Big[\bar{P}_{2n}^{2m}(\cos \theta)\cos(2m\phi)g_E(\phi_a,\theta_a)\Big] \right.\right.\\ &\left.\left.+\sin(2m\hat{\zeta}_{b,b'})\mathbb{E}\Big[\bar{P}_{2n}^{2m}(\cos \theta)\sin(2m\phi)g_E(\phi_a,\theta_a)\Big]\right)\right)\\
 &-\sum_{n=1}^{\infty} 4\imath(-1)^{n} j_{2n-1}\big(\frac{2 \pi}{\lambda} r\hat{c}_{b,b'}\big)  \left(\sum_{m=1}^{n}\bar{P}_{2n-1}^{2m-1}(0)\left(\cos((2m-1)\hat{\zeta}_{b,b'}) \mathbb{E}[\bar{P}_{2n-1}^{2m-1}(\cos \theta)\cos((2m-1)\phi)g_E(\phi_a,\theta_a)]
 \right. \right.\\
 &\left.\left. +\sin((2m-1)\hat{\zeta}_{b,b'})\mathbb{E}[\bar{P}_{2n-1}^{2m-1}(\cos \theta)\sin((2m-1)\phi)g_E(\phi_a,\theta_a)]\right)\right)
 \end{aligned}
 \end{equation}
 \end{figure*}

\begin{figure*}[h!]
\begin{equation}
\label{eq: rhob3}
\begin{aligned}
&\rho_b(b, b^{\prime})
 =\pi^2\tilde{a}(0,0)j_0(\frac{2 \pi}{\lambda} r\hat{c}_{b,b'})+ \sum_{n=1}^{\infty} (-1)^{n}(4n+1) j_{2n}\big(\frac{2 \pi}{\lambda} r\hat{c}_{b,b'}\big)P_{2n}(0)\pi^2 \sum_{k=-n}^n p_{n-k} p_{n+k} \tilde{a}(2k,0) \\
&+ \sum_{n=1}^{\infty} 4(-1)^{n} j_{2n}\big(\frac{2 \pi}{\lambda} r\hat{c}_{b,b'} \big)\left(\sum_{m=1}^{n}\bar{P}_{2n}^{2m}(0)\pi^2 \left(\cos(2m\zeta_{b,b'})\sum_{k=0}^n c_{2n,2k}^{2m} \tilde{a}(2k,2m) +\sin(2m\zeta_{b,b'})\sum_{k=0}^n c_{2n,2k}^{2m} \tilde{c}(2k,2m)\right)\right)\\
&-\sum_{n=1}^{\infty} 4\imath(-1)^{n} j_{2n-1}\big(\frac{2 \pi}{\lambda} r\hat{c}_{b,b'}\big)  \left(\sum_{m=1}^{n}\bar{P}_{2n-1}^{2m-1}(0)\pi^2 \left(\cos((2m-1)\zeta_{b,b'})\sum_{k=1}^n d_{2 n-1,2 k-1}^{2 m-1} \tilde{d}(2k-1,2m-1)
\right. \right.\\
&\left.\left. +\sin((2m-1)\zeta_{b,b'})\sum_{k=1}^n d_{2 n-1,2 k-1}^{2 m-1} \tilde{b}(2k-1,2m-1) \right)\right)
\end{aligned}
\end{equation}
\end{figure*}

\subsection{Derivation of the channel spatial correlation }
With the radiation pattern correlation now established, we can derive a closed-form expression for the spatial correlation function of the NLOS component of the channel in~\eqref{eq:nlos_component}. Assuming that the antennas located on the bottom face and the curved surface are sufficiently spatially separated to be considered uncorrelated, the spatial correlation function for user can be expressed as:
\begin{equation}
\mathbf{R} =
\begin{bmatrix}
\mathbf{R}^{\text{curve}} & \mathbf{0}_{N_hN_v \times N_b} \\
\mathbf{0}_{N_b \times N_h N_v} & \mathbf{R}^{\text{bottom}}
\end{bmatrix}
\end{equation}
where 
\begin{align}
[{\bf R}^{\rm curve}]_{(z-1)N_v+s,(z'-1)N_v+s'}&= \rho((z,s),(z',s'))\\
[\mathbf{R}^{\text{bottom}}]_{b,b'}&=\rho_b(b,b')
\end{align}
where $\rho((z,s),(z',s'))$ and $\rho_b(b,b')$ denote the radiation pattern correlation derived in \eqref{Eq:rho_final} and \eqref{eq: rhob3} taken over the azimuth and zenith distributions of cluster $n$.

\section{Numerical Results}
\label{section: simulation}
We consider a single HASP with a cylindrical MIMO system positioned at an altitude of $H=20$km. 
The antenna configuration parameters are summarized in Table~\ref{Table:antenna}. 
\begin{table}[!h]
\centering
\caption{Parameter for antenna structure}
\label{Table:antenna}
\begin{tabular}{|p{.1\textwidth} | p{.15\textwidth} | }
\hline
$N_h$ &$16$\\
  \hline
$N_v$&$6$\\
  \hline
$N_b$& $10$ \\
  \hline
$d_h,d_v,d_b$& $0.5\lambda$\\
  \hline
$R$& $\frac{N_hd_h}{2\pi}$\\
  \hline
$r$& $\frac{N_bd_b}{2\pi}$\\
 \hline
$\theta_{3dB}$& $65^{\circ}$\\
  \hline
$\phi_{3dB}$& $65^{\circ}$\\
  \hline
$G_{E,max}^{\mathrm{Tx}}$& $8$dBi\\
  \hline
\end{tabular}
\end{table}
%The numerical results validate the derivation of the SCF for a single user. 
In the sequel, we present a set of numerical results to validate the derivation of the spatial correlation function and to provide useful insights into the impact of the angular distribution and antenna geometry on the resulting correlation behavior.

\subsection{Validation of the SCFs}
For the purpose of validation, 
the zenith angles are generated according to the Laplacian density spectrum, with mean AoD $\theta_0$ and spread $\sigma$.
The corresponding probability density function is given by
\begin{equation}
f_\theta(\theta) \propto \exp \left(-\frac{\sqrt{2}\left|\theta-\theta_0\right|}{\sigma}\right) \sin \theta.
\end{equation}
The density function decays exponentially
and is zero for $\theta \notin[0, \pi]$. The normalization constant $A$ is determined by enforcing the condition that the integral of the probability density function over its support must equal unity:
\begin{equation}
\int_0^\pi A \exp \left(-\frac{\sqrt{2}\left|\theta-\theta_0\right|}{\sigma}\right) \sin \theta d \theta=1
\end{equation}
Solving this normalization integral yields the explicit expression for the normalization constant:
\begin{equation}
A=\frac{2+\sigma^2}{2 \sqrt{2} \sigma \sin \theta_0+2 \sigma^2 e^{-\frac{\pi}{\sqrt{2} \sigma}} \cosh \left(\frac{\sqrt{2}\left(\frac{\pi}{2}-\theta_0\right)}{\sigma}\right)} .
\end{equation}
The azimuth angles are generated according to a von Mises distribution with mean direction $\mu$ and concentration parameter $\kappa$, where larger values of $\kappa$ correspond to a narrower angular spread around $\mu$. The corresponding pdf is given by
\begin{equation}
f_\phi(\phi)=\frac{\exp\!\big(\kappa \cos (\phi-\mu)\big)}{2 \pi I_0(\kappa)} .
\end{equation}
where $I_0(\kappa)$ denotes the modified Bessel function of the first kind and order zero.

\par
The validation of the theoretical results computed using \eqref{Eq:rho_final} and \eqref{eq: rhob3} is done by comparison with Monte-Carlo simulation results.
The Monte-Carlo simulations are performed over $100000$
channel realizations to compute the correlation values \eqref{eq:SCF_cyl} and \eqref{eq:SCF_bot}. 
In the simulation, $\theta_0 = \frac{2\pi}{3}$, $N_0 = 14$, $\sigma = 15^\circ$, $\kappa = 10$, and $\mu = \frac{\pi}{6}$ are used to validate the antenna elements on the curved surface. For the bottom antenna elements, $\theta_0 = \frac{11\pi}{12}$ is adopted to validate their correlation, as these elements are intended to serve users directly beneath the HAPS.
\par
In order to remove the effect of the antenna gains, we normalize the spatial correlation function by the square root of the corresponding self-correlations. Specifically, for two antenna elements located at $(z,s)$ and $(z',s')$ on the curved surface, we define the normalized correlation coefficient as
\begin{equation}
\overline{\rho}\big((z,s),(z',s')\big)
:=\frac{\rho\big((z,s),(z',s')\big)}
{\sqrt{\rho\big((z,s),(z,s)\big)\,\rho\big((z',s'),(z',s')\big)}}.
\end{equation}
Similarly, for two elements $b$ and $b'$ on the bottom surface, we define
\begin{equation}
\overline{\rho}_b({b,b'})
:=\frac{\rho_b(b,b')}
{\sqrt{\rho_b(b,b)\,\rho_b(b',b')}}.
\end{equation}
%\textcolor{green}{For notational clarity, $\bar{\rho}$ is used to represent the normalized correlation among antenna elements on the curved surface, whereas $\bar{\rho}_b$ denotes the normalized correlation for the antenna elements located at the bottom.}
Throughout the simulations, each antenna element on the curved surface is indexed by $(z,s)$, where $z$ denotes the $z^{\text{th}}$ element in the $s^{\text{th}}$ column, as shown in Fig.~\ref{Fig: AN_Tx}. The element $(1,1)$ is selected as the reference element, and the spatial correlation between this reference element and the other antenna elements is evaluated.
The antenna elements on the bottom ring are indexed sequentially, as illustrated in Fig.~\ref{Fig: AN_Tx_circular}. Element $1$ is chosen as the reference element.
\subsubsection{Impact of Azimuth Angular Distribution on Spatial Correlation}
\begin{figure*}[h]
\centering
\subfigure[Correlation for elements in the same ring with reference to element $(1,1)$]{
    \includegraphics[width=3in]{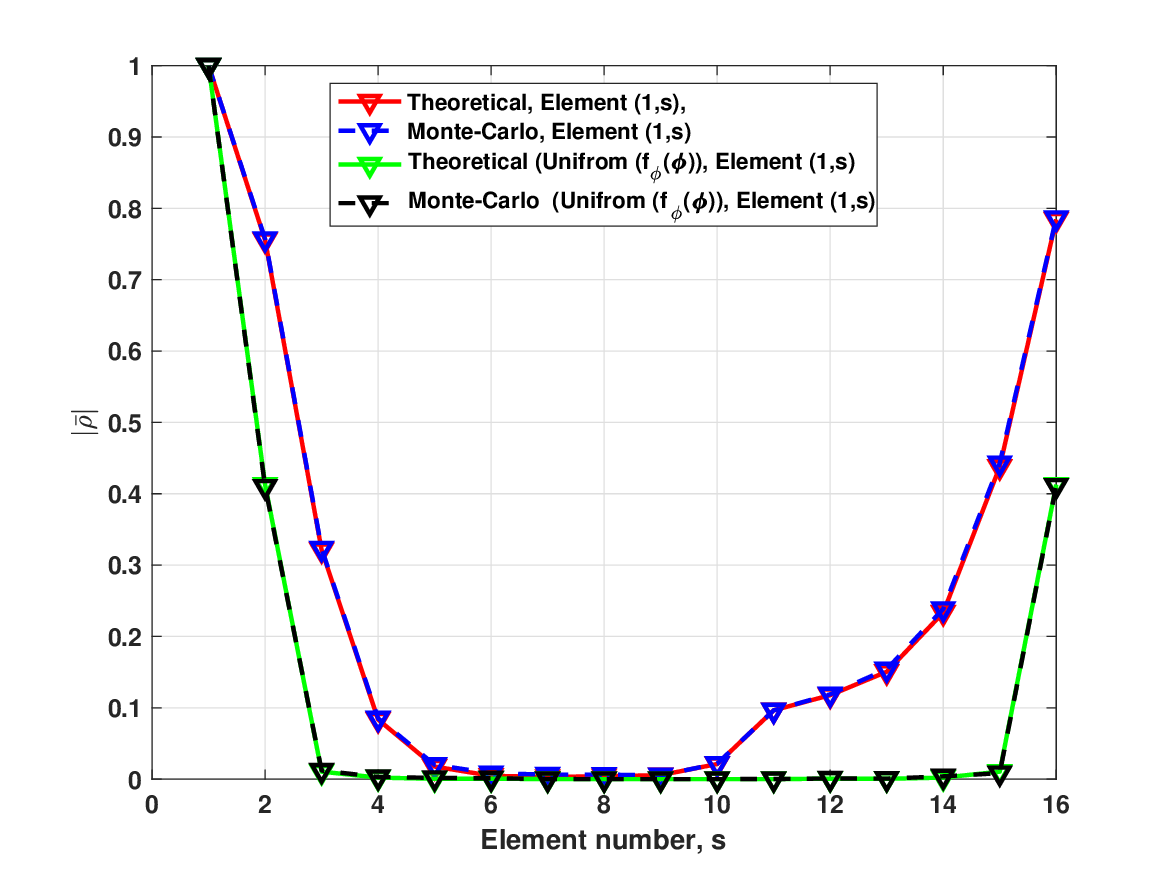}
    \label{Fig:circular_corr_uniform}
}
\subfigure[Correlation for elements in the same column with reference to element $(1,1)$]{
    \includegraphics[width=3in]{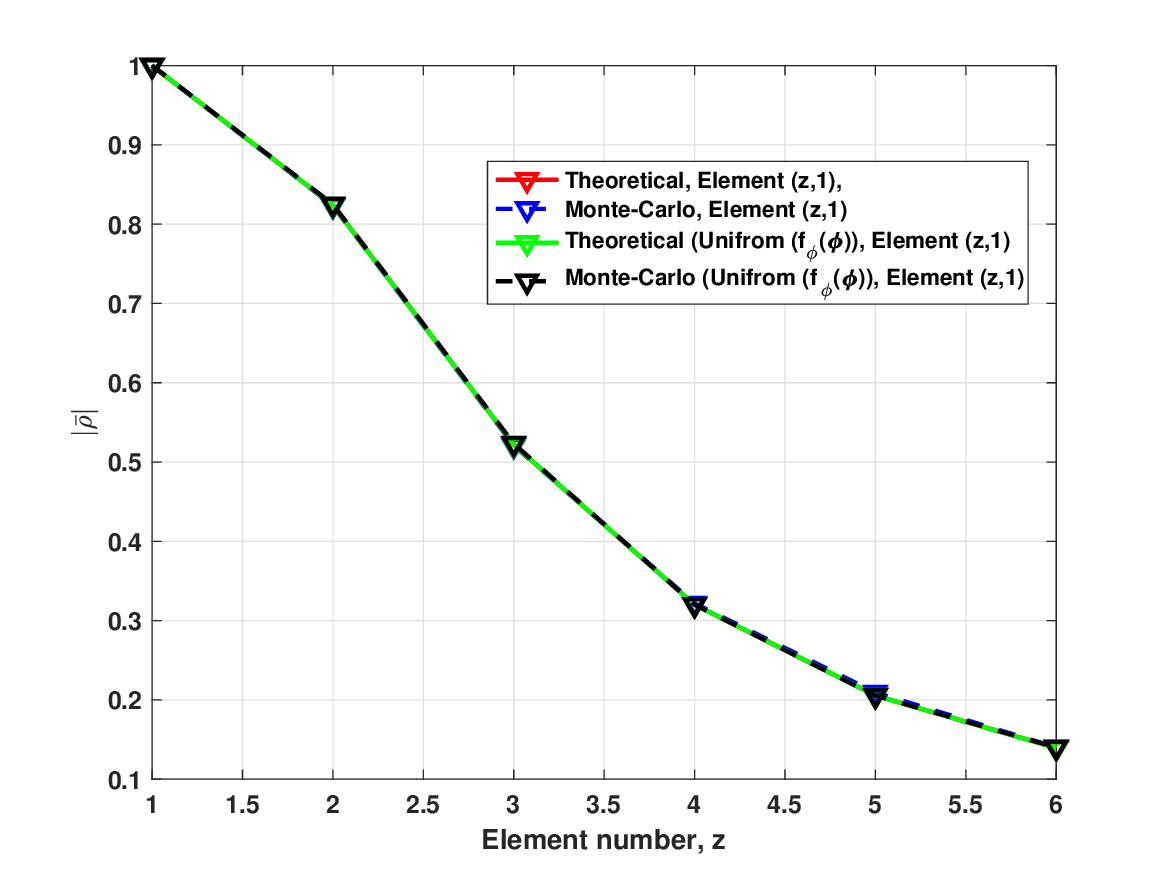}
    \label{Fig:linear_corr_uniform}
}
\caption{Correlation validation of surface cylindrical antenna array.}
\label{Fig:corr_curve_uniform}
\end{figure*}
Fig.~\ref{Fig:corr_curve_uniform} illustrates the correlation coefficients between the antenna element $(z,s)$ and the reference element $(1,1)$, under both von Mises and uniform azimuth angular distributions with support in the interval $[-\frac{5\pi}{6},\,\pi]$.
The results demonstrate that the theoretically derived correlations agree well with the Monte Carlo simulation results for both angular distributions, indicating that the proposed method is applicable to arbitrary underlying angular distributions.
Specifically, Fig.~\ref{Fig:circular_corr_uniform} presents the correlation of elements located on the first ring with respect to the reference element $(1,1)$. It can be observed that the correlation initially decreases and then increases. This behavior arises because, as the element index along the circular ring increases, the distance between the element and $(1,1)$ first increases and subsequently decreases due to the circular geometry.
Furthermore, the correlation values corresponding to the uniform angular distribution are generally lower than those of the Von Mises distribution. This can be explained by noting that the uniform distribution has an angular variance of $25\pi^2/108$, which is significantly larger than the angular spread assumed for the Von Mises distribution, leading to reduced spatial correlation.
\par
In addition, Fig.~\ref{Fig:linear_corr_uniform} presents the correlation coefficients of antenna elements located in the same vertical column with respect to the reference element $(1,1)$.
In addition, Fig.~\ref{Fig:linear_corr_uniform} shows that the correlation monotonically decreases with increasing $z$. This trend is attributed to the fact that, within the same column, the distance from the reference element $(1,1)$ increases as $z$ increases.
Moreover, the correlation values obtained under the uniform azimuth angular distribution coincide with those under the von Mises distribution. This can be explained by the fact that antenna elements located in the same column share an identical azimuth angle, which eliminates the impact of the azimuth angular distribution on the correlation.
Specifically, the correlation between elements $(z,s)$ and $(z,s')$ is given by
\[
\rho((z,s),(z',s'))
=
\mathbb{E}\big[
g_{E,H}^{s,s}(\phi)\,
g_{E,V}(\theta)\,
\exp(\mathrm{i} k d_v (z-z')\cos\theta)
\big].
\]
Since the azimuth angle $\phi$ and the zenith angle $\theta$ are statistically independent, the normalized correlation reduces to
\[
\bar{\rho}((z,s),(z',s'))
=
\frac{
\mathbb{E}\big[
g_{E,V}(\theta)\,
\exp(\mathrm{i} k d_v (z-z')\cos\theta)
\big]
}{
\mathbb{E}\big[
g_{E,V}(\theta)
\big]
},
\]
which is independent of the azimuth angular distribution.

\begin{figure}[h!]
\centering
\includegraphics[width=3in]{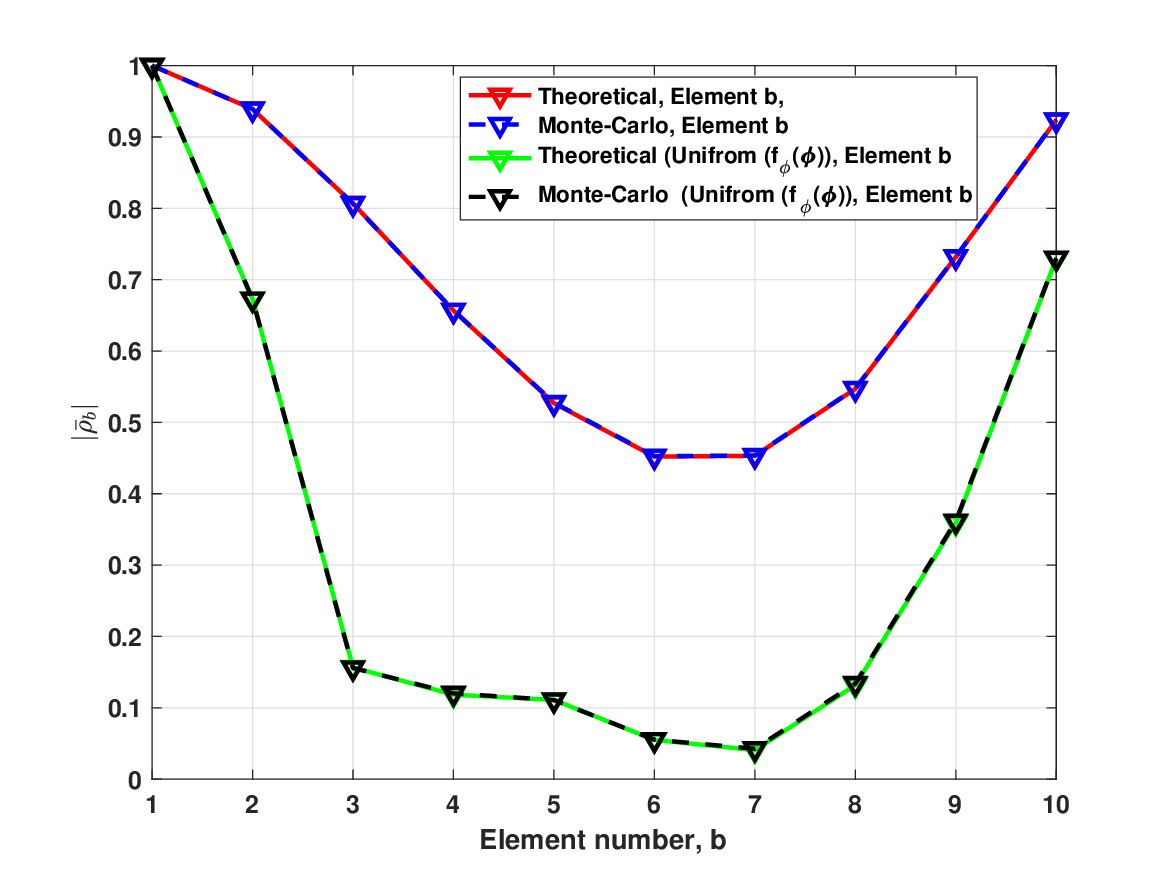}
\caption{Correlation for elements in the bottom ring with reference to element $1$.}
\label{Fig:corr_bottom_uniform}
\end{figure}
We next present in Fig.~\ref{Fig:corr_bottom_uniform} the correlation values of the 
$b$-th element in the circular antenna array at the bottom, with respect to the reference element.
It is also evident that the correlation values associated with the uniform distribution are generally lower.
For both angular distributions, the correlation initially decreases and then increases as the element index grows, which can be attributed to the circular geometry of the antenna array.

\subsubsection{Impact of Antenna Spacing on Spatial Correlation}
\begin{figure}[h!]
\centering
\includegraphics[width=3in]{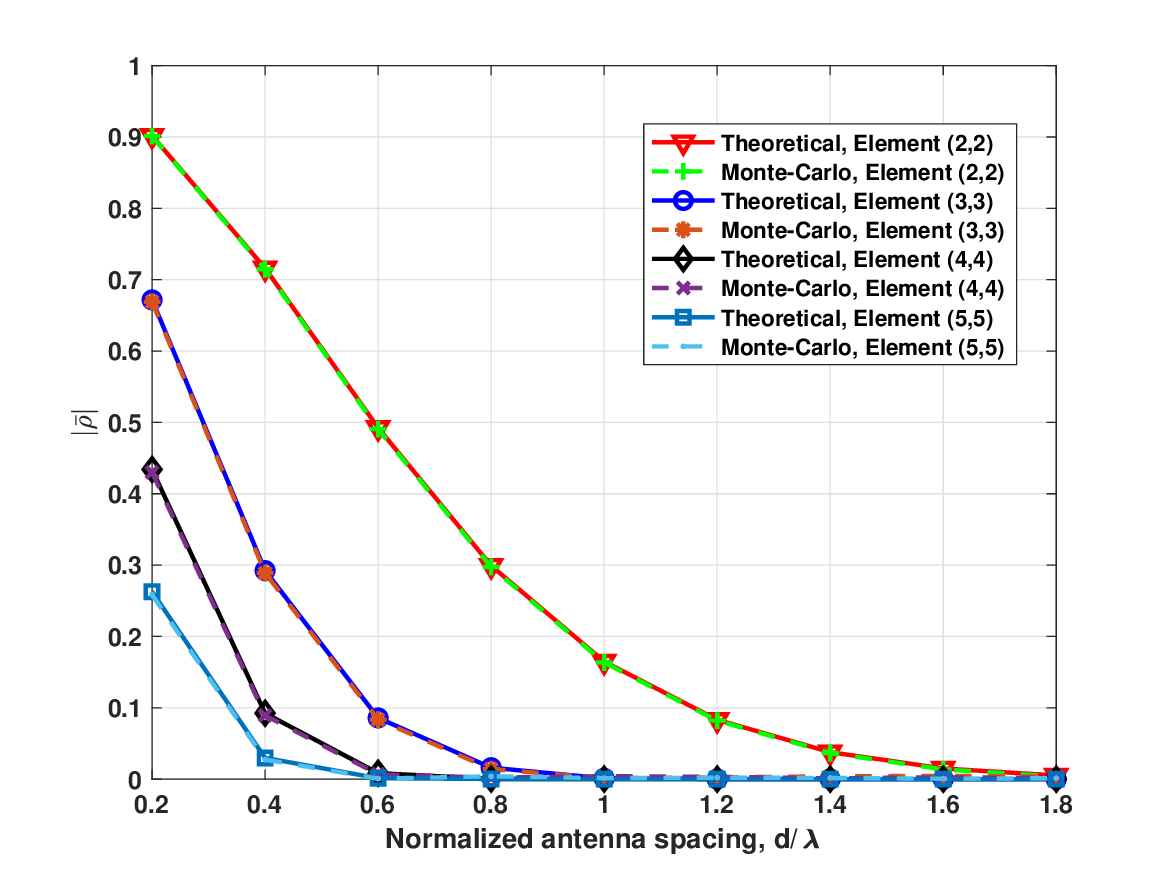}
\caption{Correlation with respect to element $(1,1)$ on the curved surface.}
\label{Fig:curve_d}
\end{figure}
Fig.~\ref{Fig:curve_d} depicts the correlation between elements $(2,2)$, $(3,3)$, $(4,4)$, and $(5,5)$ and the reference element $(1,1)$ on the curved surface as a function of $d/\lambda$. For notational simplicity, we set $d = d_v = d_h = d_b$ in the following simulations.
As expected, the correlation values decrease with increasing antenna spacing. In addition, as the distance between the considered element and the reference element $(1,1)$ increases, the corresponding correlation decreases, i.e., the correlation progressively reduces from elements $(2,2)$ to $(5,5)$. %Moreover, the theoretical results exhibit excellent agreement with the Monte Carlo simulation results.
\begin{figure}[h!]
\centering
\includegraphics[width=3in]{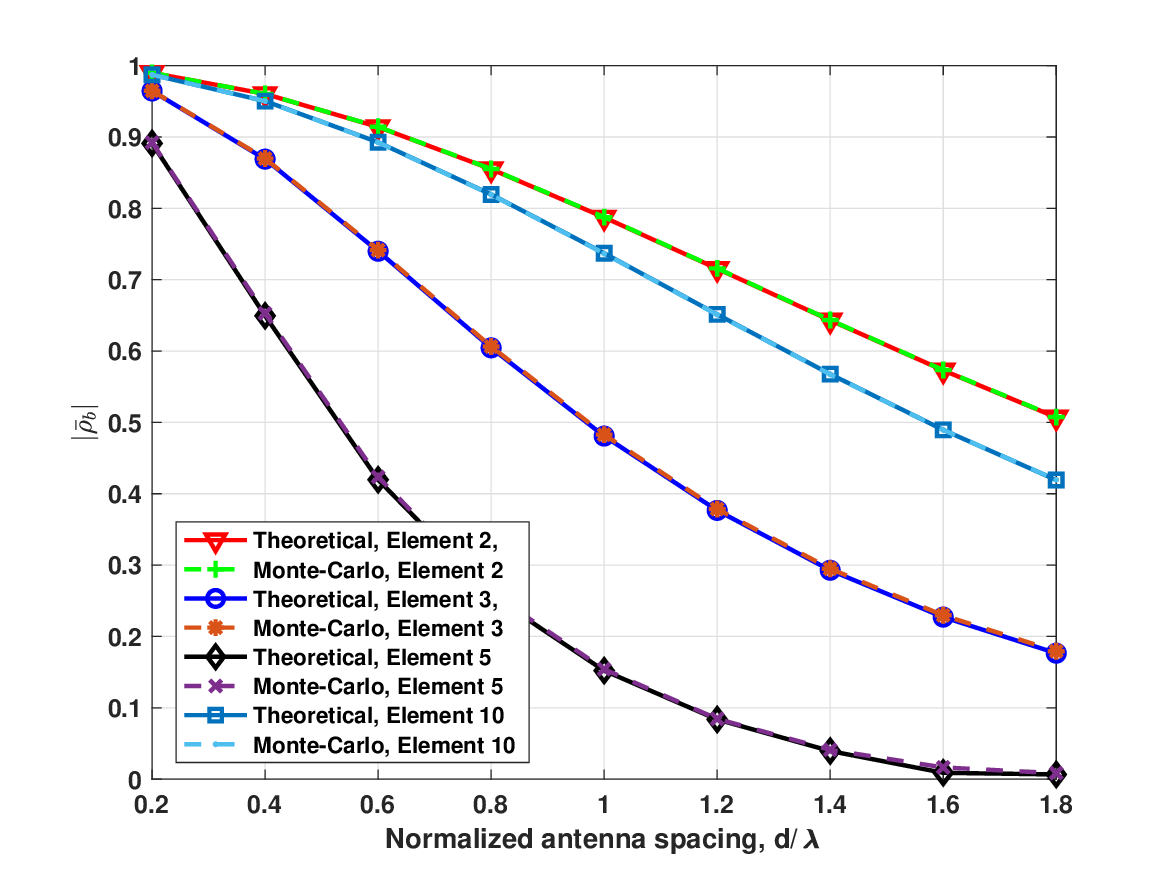}
\caption{Correlation with respect to element $1$ on the bottom ring.}
\label{Fig:bottom_d}
\end{figure}
\par
Fig.~\ref{Fig:bottom_d} illustrates the correlation between elements $2$, $3$, $5$, and $10$ and the reference element $1$ in the bottom circular antenna array as a function of $d/\lambda$. 
As expected, the correlation values decrease as the antenna spacing increases. Moreover, due to the circular geometry of the array, the correlation values differ among the elements even at the same spacing. Specifically, the correlation follows the order of elements $2$, $10$, $3$, and $5$, which can be attributed to their different spatial distances from the reference element $1$.
\subsubsection{Impact of Azimuth Angular Spread on Spatial Correlation}
\begin{figure}[h!]
\centering
\includegraphics[width=3in]{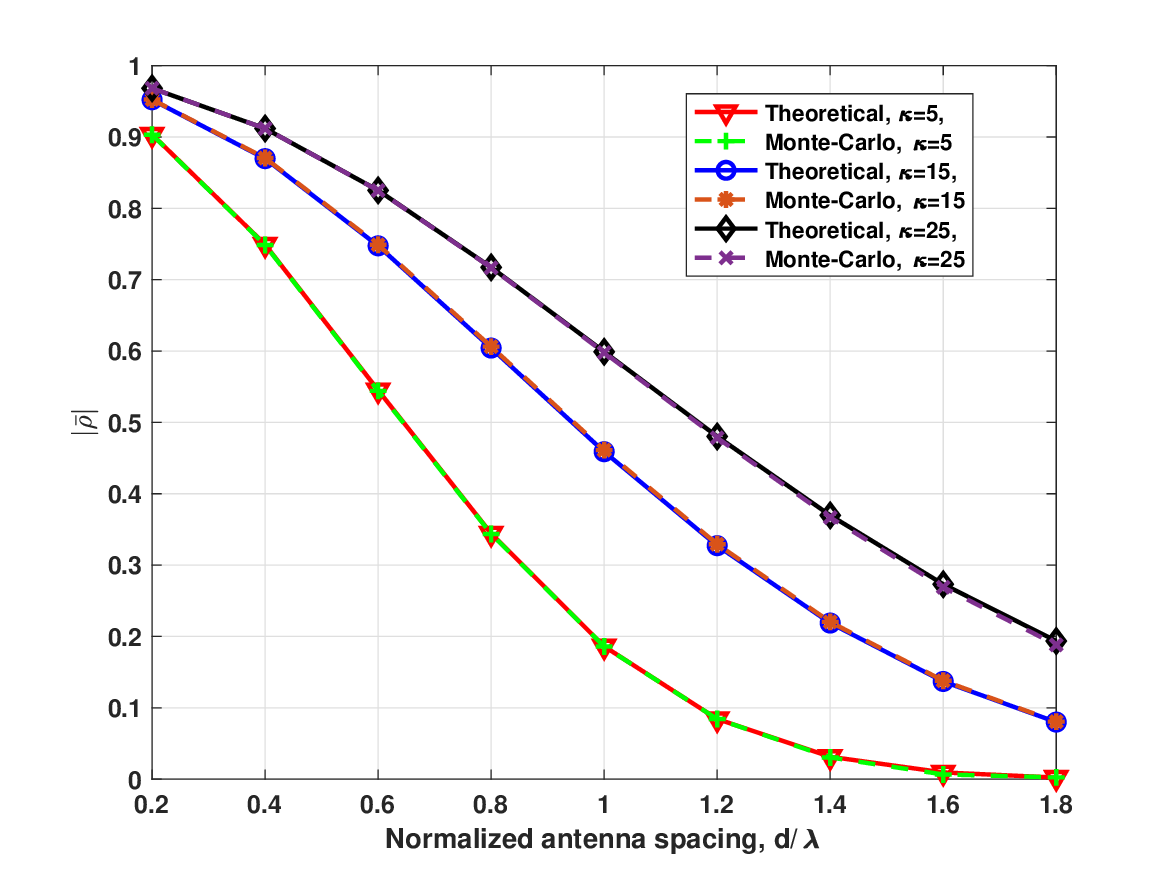}
\caption{Effect of azimuth angular spread on correlation over the curved surface.}
\label{Fig:curve_kappa}
\end{figure}
To illustrate the impact of the azimuth angular spread on spatial correlation, Fig.~\ref{Fig:curve_kappa} presents the correlation between elements $(1,1)$ and $(1,2)$ as a function of $d/\lambda$ for different values of the azimuth angular spread parameter $\kappa$ on the curved surface. It can be observed that larger values of $\kappa$ result in higher correlation levels. This is because an increase in $\kappa$ corresponds to a more concentrated azimuth angular distribution, i.e., a smaller angular spread, which reduces the phase dispersion across the antenna elements and thereby enhances the spatial correlation. In addition, the correlation values decrease as the antenna spacing increases.% and an excellent agreement is observed between the theoretical results and the Monte Carlo simulations.
\par
\begin{figure}[h!]
\centering
\includegraphics[width=3in]{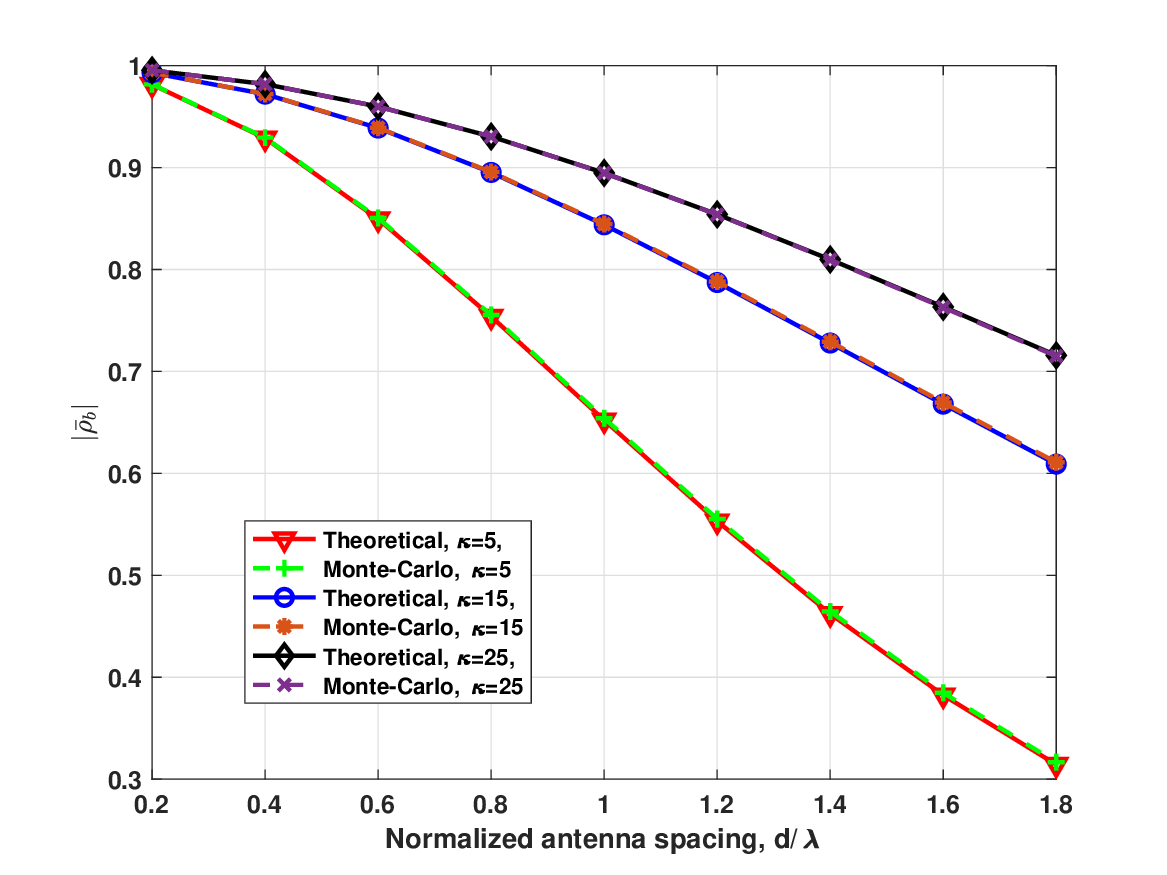}
\caption{Effect of azimuth angular spread on correlation over the bottom ring.}
\label{Fig:bottom_kappa}
\end{figure}
Fig.~\ref{Fig:bottom_kappa} further investigates the correlation between element $2$ and element $1$ on the bottom ring as a function of $d/\lambda$ for different values of the azimuth angular spread parameter $\kappa$. It is observed that increasing $\kappa$ leads to higher correlation values owing to the reduced angular spread. In addition,  the correlation values decrease as the antenna spacing increases.

\subsubsection{Impact of Zenith Angular Spread on Spatial Correlation}
\begin{figure}[h!]
\centering
\includegraphics[width=3in]{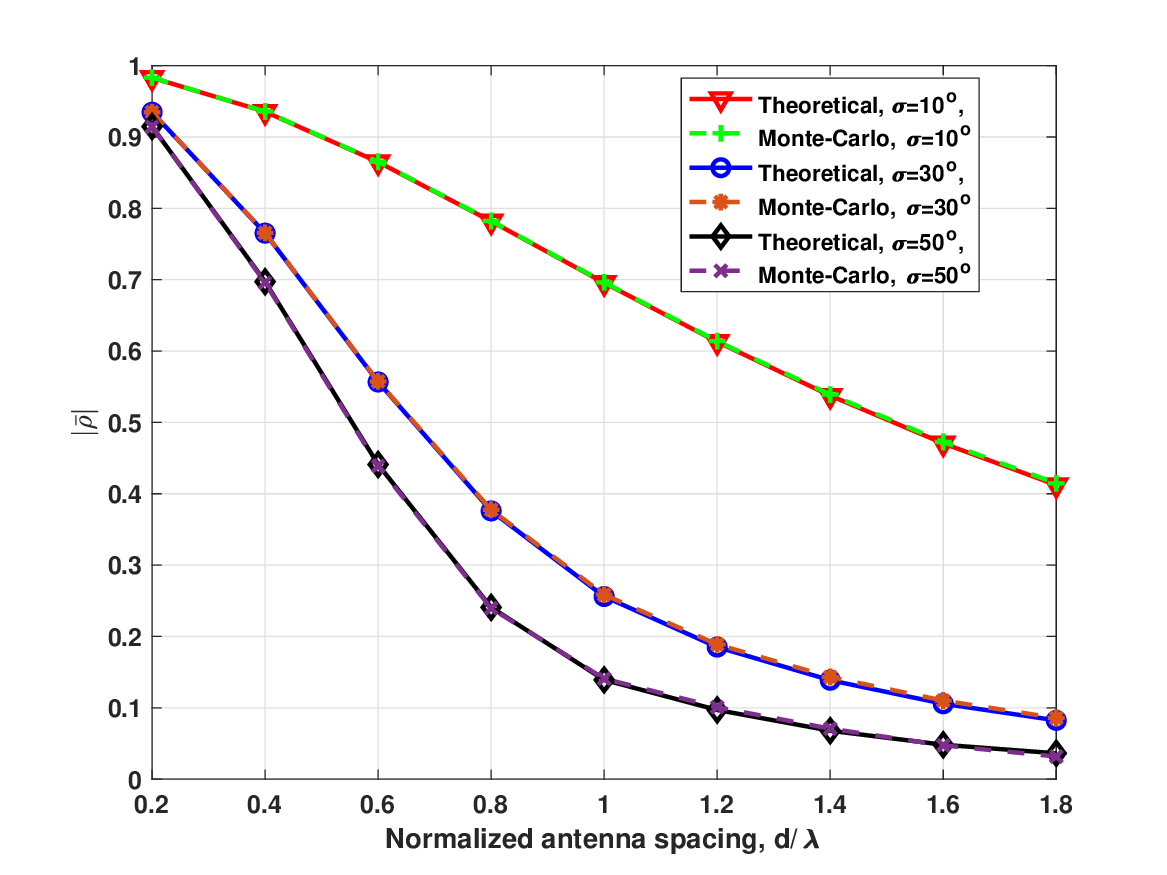}
\caption{Effect of zenith angular spread on correlation over the curved surface.}
\label{Fig:curve_sigma}
\end{figure}
Fig.~\ref{Fig:curve_sigma} illustrates the correlation between elements $(1,1)$ and $(2,1)$ as a function of $d/\lambda$ for different values of the zenith angular spread parameter $\sigma$ on the curved surface. The correlation values decrease with increasing antenna spacing. Moreover, for a fixed antenna spacing, the correlation decreases as the zenith angular spread $\sigma$ increases, since a wider zenith angular distribution introduces greater phase dispersion across the antenna elements, thereby reducing the spatial correlation.
\par
\begin{figure}[h!]
\centering
\includegraphics[width=3in]{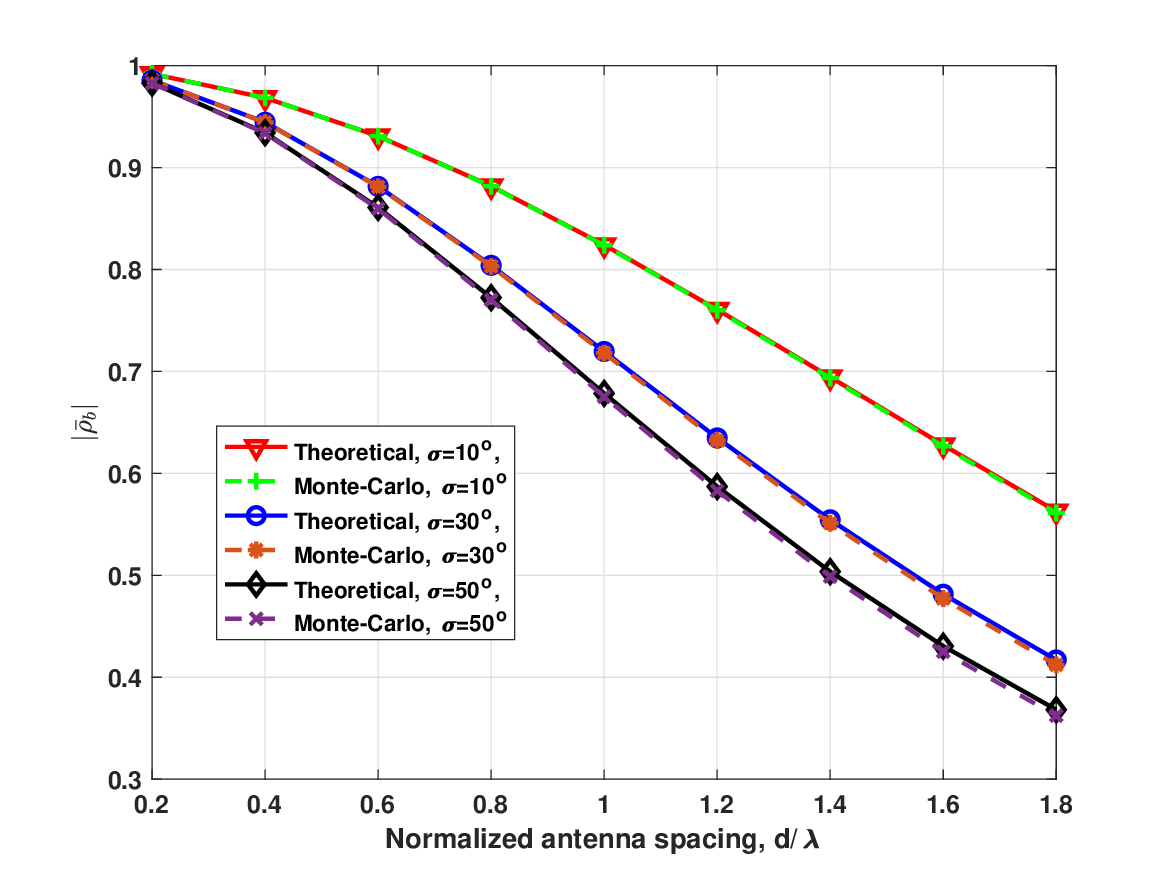}
\caption{Effect of zenith angular spread on correlation over the bottom ring.}
\label{Fig:bottom_sigma}
\end{figure}
Fig.~\ref{Fig:bottom_sigma} shows the correlation between elements $2$ and $1$ on the bottom ring versus $d/\lambda$ for different zenith angular spreads $\sigma$. The correlation decreases with increasing $\sigma$, and the theoretical results closely match the Monte Carlo simulations.
\par
% \begin{figure}[!h]
% \centering
% \includegraphics[width=5in]{Fig_R_heatmap.pdf}
% \caption{Normalized Spatial Correlation Matrix $\mathbf{R}$}
% \label{Fig:R_heatmap}
% \end{figure}
\begin{figure}[!h]
\centering
\includegraphics[width=5in,trim=4cm 0cm 0cm 0cm,clip]{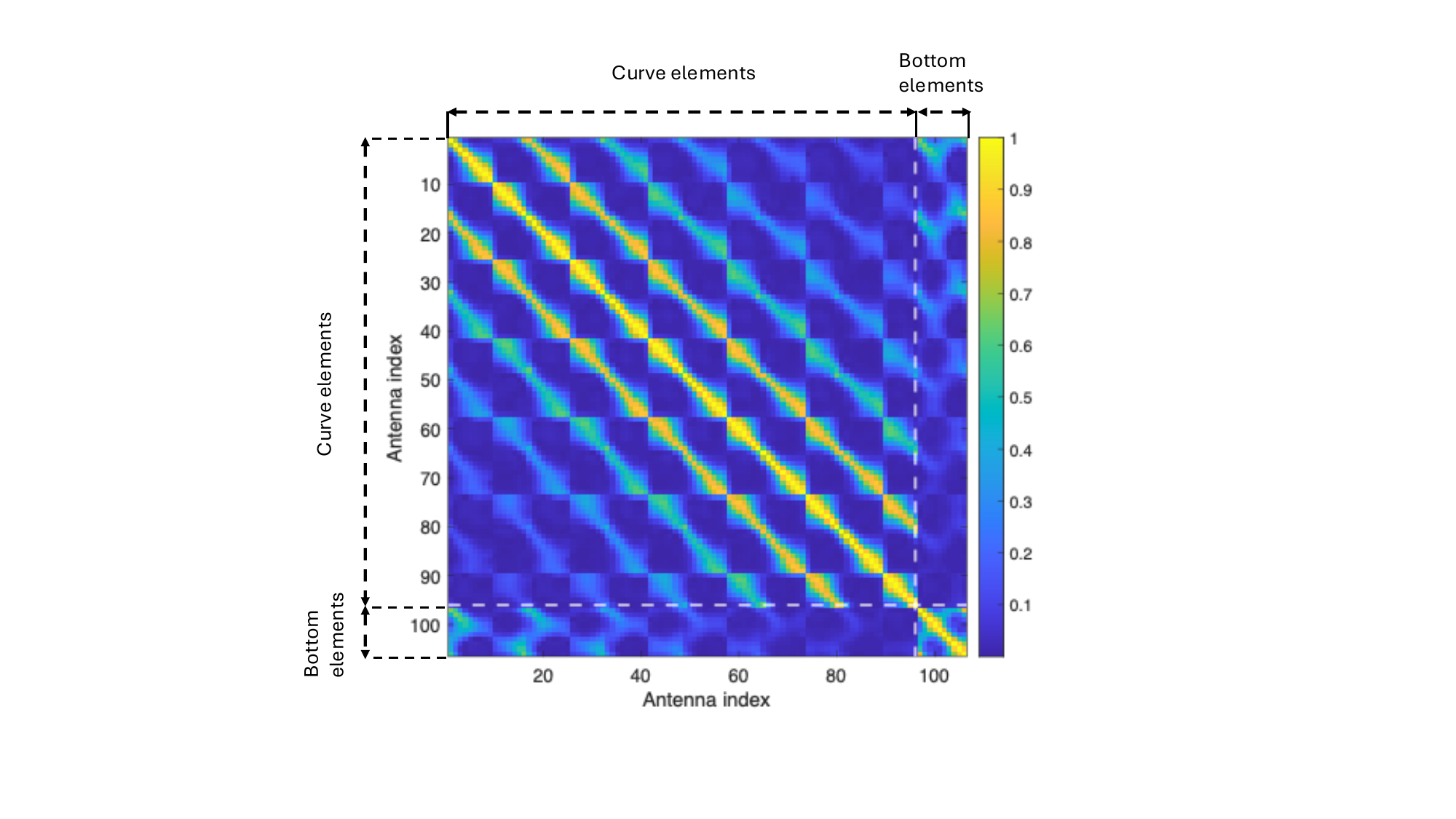}
\caption{Normalized Spatial Correlation Matrix $\mathbf{R}$}
\label{Fig:R_heatmap}
\end{figure}
Fig. \ref{Fig:R_heatmap} shows the normalized spatial correlation matrix obtained directly from Monte Carlo channel realizations. Specifically, the correlation matrix is computed as
$
\mathbb{E}\!\left[{\mathbf h}^{\mathrm{NLOS}}
{(\mathbf h}^{\mathrm{NLOS}})^{H}\right]
$.
The upper-left block corresponds to the correlation matrix of the curved-surface antenna elements, while the lower-right block corresponds to the correlation matrix of the bottom antenna elements. The upper-right and lower-left blocks represent the cross-correlations between the curved-surface and bottom antenna groups.
\par
It can be observed that the cross-correlation blocks exhibit significantly lower correlation levels than the two diagonal blocks. This is because the boresight directions of the curved-surface and bottom antenna elements are nearly orthogonal, resulting in a substantial reduction in their common angular visibility. Consequently, the overlap of their antenna radiation patterns is limited, which suppresses the contribution of common scatterers to the spatial correlation.
The numerical results therefore, confirm that the correlations between the curved-surface and bottom antenna elements are much weaker than the intra-group correlations. This observation supports the use of a block-diagonal approximation for the spatial correlation matrix.
\par
\begin{figure}[!h]
\centering
\includegraphics[width=3in]{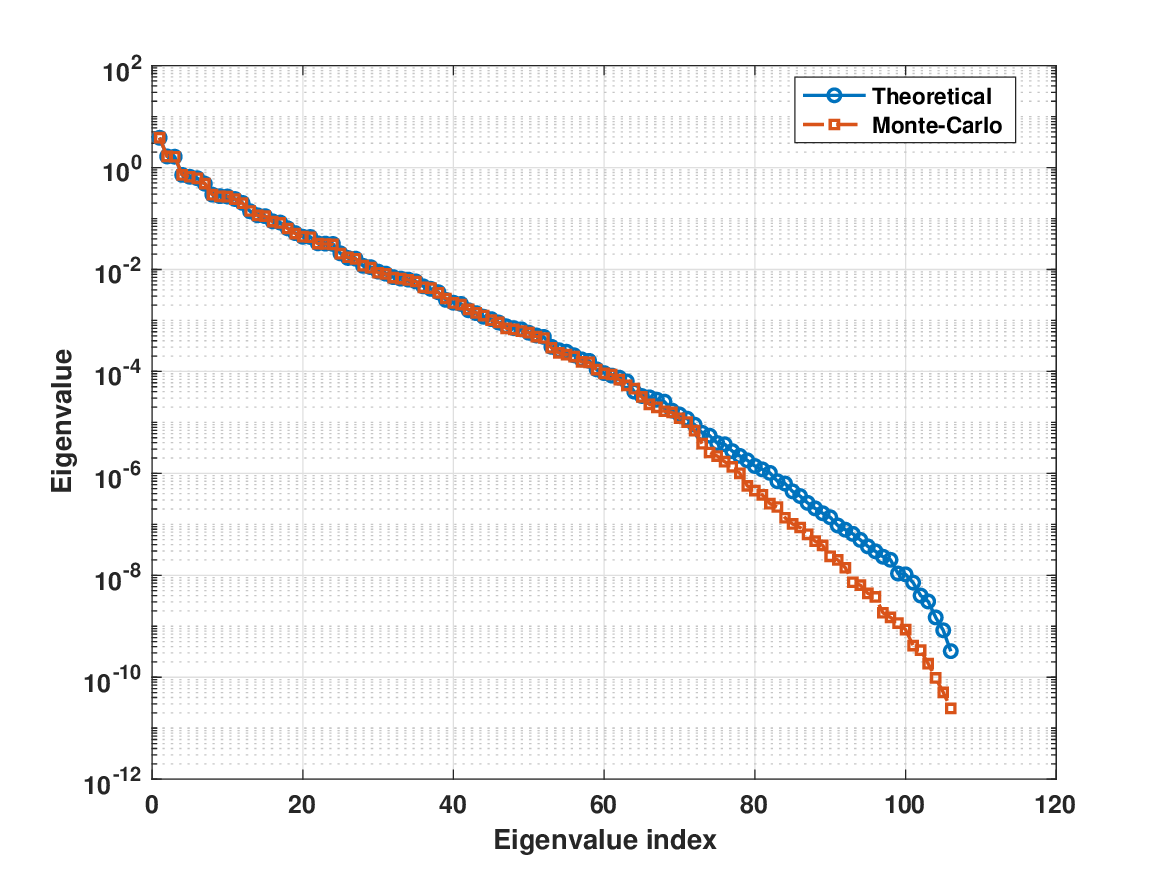}
\caption{Eigenvalue spectrum of the  correlation matrix $\mathbf{R}$}
\label{Fig:eig}
\end{figure}
To further validate the proposed closed-form SCF at the matrix level, we compare the eigenvalue distribution of the spatial correlation matrix $\mathbf R_{\rm Tx}$ obtained from the analytical SCF with that of the Monte Carlo benchmark $\mathbf R_{\rm Tx,MC}$. As shown in Fig.~\ref{Fig:eig}, the two eigenvalue distribution exhibit excellent agreement over several orders of magnitude, with nearly identical dominant eigenvalues. This demonstrates that the proposed analytical SCF accurately captures the principal spatial correlation structure and the effective spatial degrees of freedom of the antenna array.
In addition, we perform a two-sample Kolmogorov--Smirnov (K--S) test on the normalized eigenvalue sets. The resulting K--S statistic is
\begin{equation}
D_{\rm KS}=0.075472,
\end{equation}
indicating a high degree of consistency between the eigenvalue distributions obtained from the proposed analytical SCF and the Monte Carlo benchmark.

\subsection{Performance Evaluation Using the Proposed SCF}
To investigate the impact of the proposed SCF on practical massive MIMO systems, we consider a downlink multi-user MISO scenario, where a HAPS equipped with $M$ antennas serves $K$ single-antenna users. The HAPS operates at an altitude of $20$ km, and the users are uniformly distributed within a circular coverage area with radius $R_{\mathrm{radius}}$ km. For user $l$ located at $(x_l,y_l,z_l)$, the corresponding LoS zenith and azimuth angles are given by
$
\theta_l^{LoS}=\frac{\pi}{2}+\arctan\left(\frac{H-z_l}{\sqrt{x_l^2+y_l^2}}\right),
$
and 
$
\phi_l^{LoS}=\arctan\left(\frac{x_l}{y_l}\right),
$
where $H$ denotes the altitude of the HAPS.
\par
\subsubsection{Kronecker Channel Model}
Based on the above correlation structure, the complete Kronecker channel between the HAPS and user $l$ is modeled using a Rician fading model as:
\begin{equation}
\label{Eq:channel}
\mathbf{h}_l=\sqrt{\beta}(\sqrt{\frac{\kappa_l}{\kappa_l+1}}\mathbf{h}_l^{\mathrm{LoS}}+\sqrt{\frac{1}{\kappa_l+1}} \mathbf{h}_l^{\mathrm{NLoS}},
\end{equation}
where $\kappa_l$ is the Rician K-factor, ${\mathbf{h}}_l^{\mathrm{NLoS}}=\mathbf{R}_l^{\frac{1}{2}}\mathbf{z}$ is NLoS components with the entries of $\mathbf{z} \in \mathbb{C}^{M \times 1}$ are independently
and identically distributed according to a complex circularly
symmetric Gaussian distribution, i.e., $\mathcal{CN}(0,1)$, and
$\mathbf{h}_l^{\mathrm{LoS}}=\exp(-j2\pi\frac{d_{3D}}{\lambda})\sqrt{g_E\left(\phi_l^{LoS}, \theta_l^{LoS} \right)}\mathbf{a}_{tx}^H(\phi_l^{LoS}, \theta_l^{LoS})$ is the LoS components.
\subsubsection{ZF precoding}
To evaluate the effect of spatial correlation on beamforming performance, the HAPS employs zero-forcing beamforming (ZFBF) to simultaneously serve the $K$ users. Let
$
\mathbf{H}=[\mathbf{h}_1,\mathbf{h}_2,\ldots,\mathbf{h}_K]
\in \mathbb{C}^{M \times K}
$
denote the composite channel matrix. The normalized ZFBF matrix is given by
\begin{equation}
    \mathbf{U}= \mathbf{H}(\mathbf{H}^H\mathbf{H})^{-1}\big( \mathcal{D}\{(\mathbf{H}^{H}\mathbf{H})^{-1}\}\big)^{-\frac{1}{2}}
\end{equation}
where $\mathcal{D}{\cdot}$ extracts the diagonal elements of a matrix.
Assuming equal power allocation among all users, the SINR of user $l$ is expressed as
\begin{equation}
{\rm SINR}_l
=
\frac{
\frac{P_t}{K}
\left|
\mathbf{h}_l^{H}\mathbf{u}_l
\right|^2
}{
\frac{P_t}{K}
\sum_{i\neq l}
\left|
\mathbf{h}_l^{H}\mathbf{u}_i
\right|^2
+
\sigma^2
}.
\end{equation}
where $P_t$ denotes the HAPS transmit power and $\sigma^2$ is the noise power. The noise power in dBm is calculated as
$
\sigma_{d B m}^2=-174+10 \log (B W)+N F
$, where ${\rm BW}=20$ MHz is the system bandwidth and $NF=9$ dB is the receiver noise figure. The HAPS transmit power is set to $P_t=51.45$ dBm.
\begin{figure}[!h]
\centering
\includegraphics[width=3in]{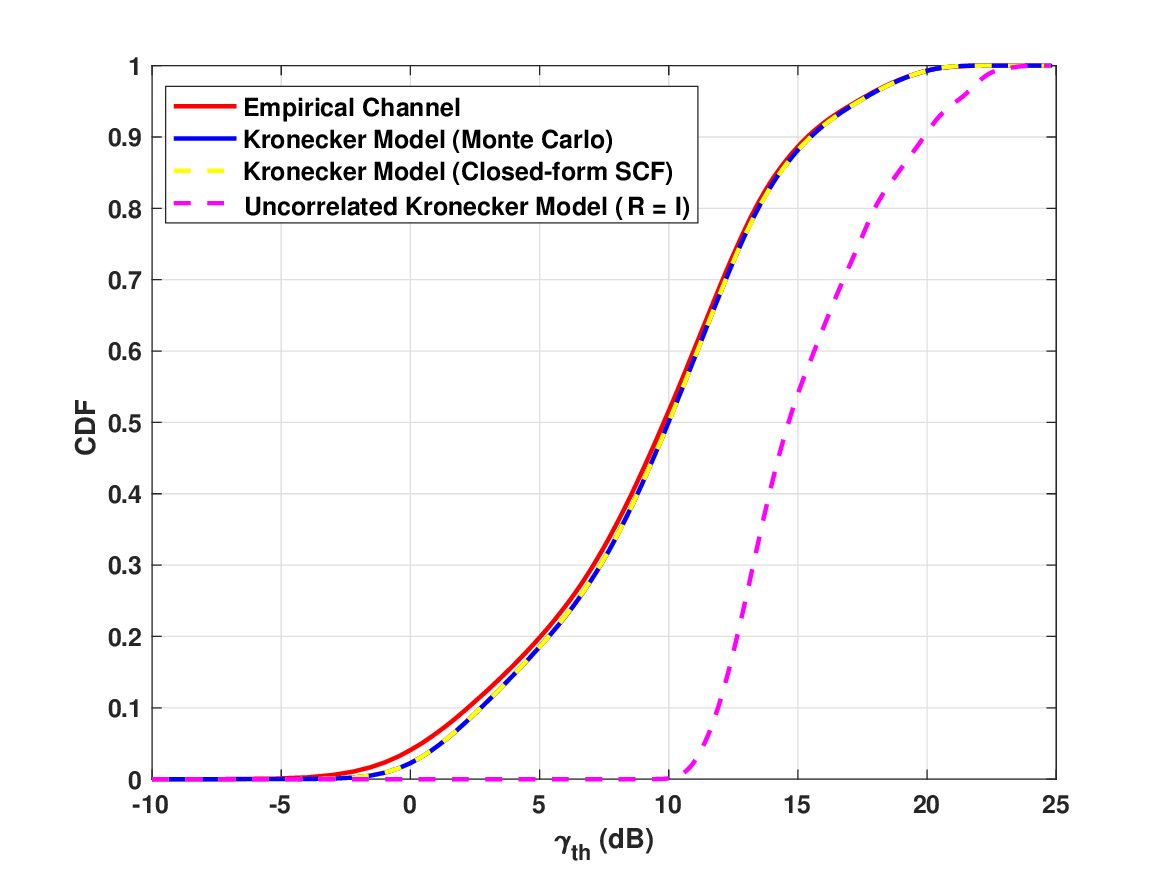}
\caption{SINR Distributions for User}
\label{Fig:SINR_CDF}
\end{figure}
\par
Fig. \ref{Fig:SINR_CDF} compares the SINR distributions obtained from the empirical channel, the proposed SCF-based Kronecker channel model, and the uncorrelated channel model under ZF beamforming. It can be observed that the SINR distributions of the empirical channel, the Monte-Carlo Kronecker model, and the closed-form SCF-based Kronecker model are nearly identical, validating the accuracy of the proposed SCF in characterizing spatial correlation. In contrast, the uncorrelated channel model (R=I) achieves significantly higher SINR, indicating that spatial correlation reduces channel orthogonality and degrades ZF beamforming performance.
\par
\begin{figure}[!h]
\centering
\includegraphics[width=3in]{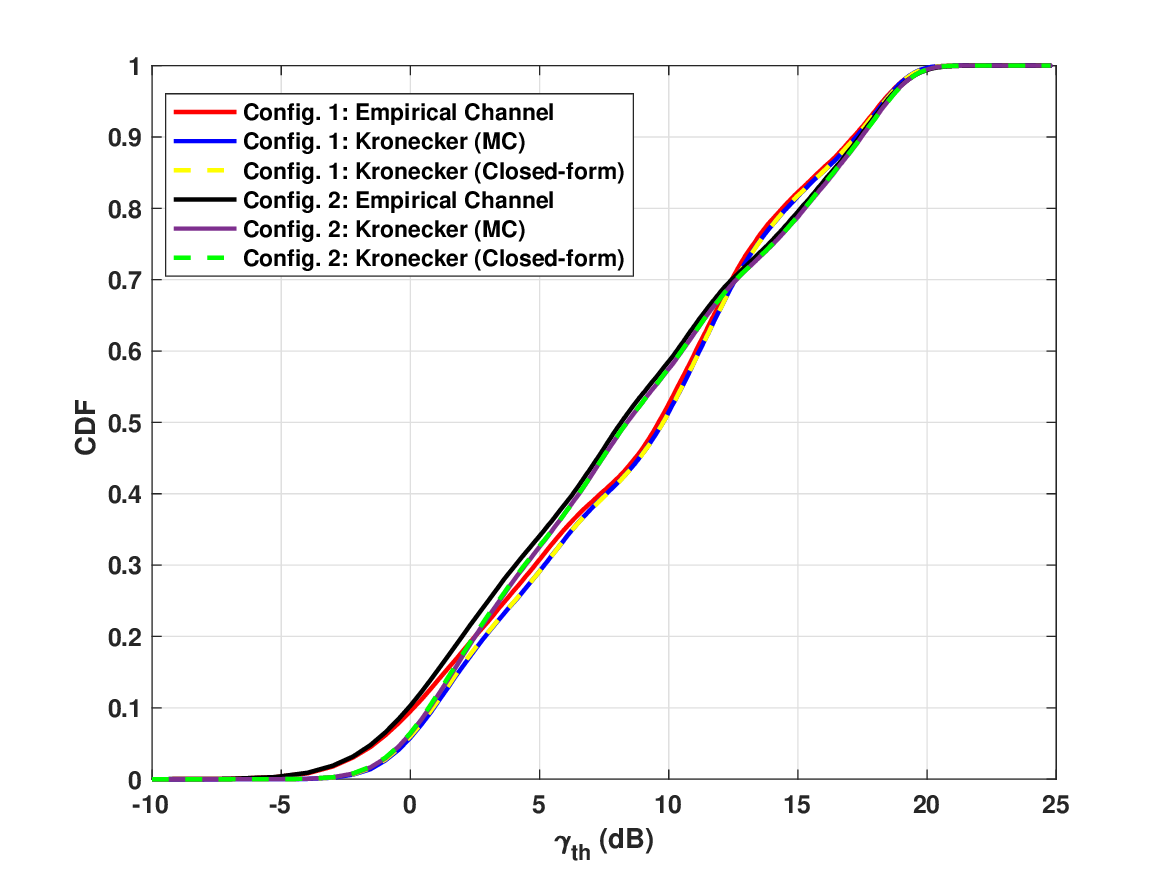}
\caption{SINR distributions for user for different antenna distribution}
\label{Fig:SINR_AN_configuration}
\end{figure}
To illustrate the impact of antenna configuration, we compare two cylindrical array deployments in Fig.~\ref{Fig:SINR_AN_configuration} while maintaining the same total number of antenna elements. Configuration~1 employs $N_h=16$ and $N_v=6$, whereas Configuration~2 employs $N_h=12$ and $N_v=8$. In both cases, the number of bottom-surface antenna elements is fixed at $N_b=10$.
\par
Fig.~\ref{Fig:SINR_AN_configuration} shows that the two antenna configurations result in different SINR distributions. Configuration~1 generally achieves better performance over most SINR thresholds due to its larger number of horizontal elements, which provides higher azimuth resolution and improves user separation under ZF beamforming. In contrast, Configuration~2 exhibits a slight advantage in the high-SINR regime, as the increased number of vertical elements offers higher elevation-domain array gain and enhances the received signal power. Consequently, Configuration~2 is capable of achieving higher SINR values for a subset of users under favorable propagation conditions.
\par
\begin{figure}[!h]
\centering
\includegraphics[width=3in]{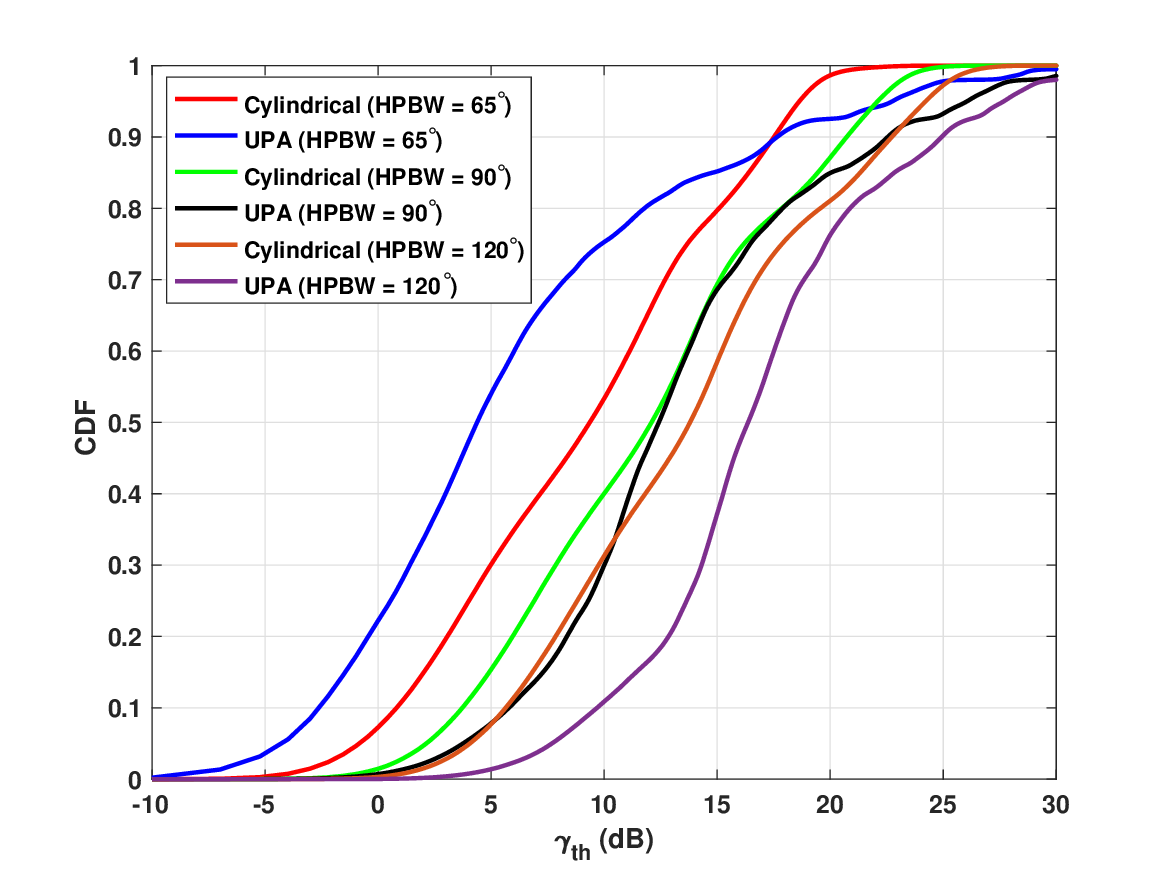}
\caption{SINR Distributions for different HPBW}
\label{Fig:HPBW}
\end{figure}
To investigate the impact of antenna directivity, we compare the proposed cylindrical array and a conventional UPA under different HPBW settings, namely $65^\circ$, $90^\circ$, and $120^\circ$, using the same 3GPP antenna pattern. 
Fig.~\ref{Fig:HPBW} shows the resulting SINR distributions. For HPBW is $65^\circ$, the cylindrical array significantly outperforms the UPA due to the directional separation provided by the different cylindrical panels. As the HPBW increases to $90^\circ$, the performance gap becomes considerably smaller. When the HPBW further increases to $120^\circ$, the cylindrical-array advantage disappears and the UPA achieves slightly better performance.
This can be explained by two factors. First, the wider antenna beam increases the overlap among the cylindrical panels, thereby reducing the geometric diversity offered by the cylindrical structure. Second, the broader radiation pattern allows the UPA to maintain a higher element gain over a wider angular region. As a result, users located at larger angular offsets from the array boresight experience improved received signal power, leading to enhanced SINR performance.

\section{Conclusion}
\label{section: conclusion}
In this paper, a closed-form expression for the SCF of three-dimensional MIMO channels with cylindrical antenna arrays deployed on HAPS is derived using the SHE of plane waves and the trigonometric expansion of the Legendre polynomials and associated Legendre polynomials. The proposed formulation accommodates arbitrary azimuth and zenith angular distributions as well as general antenna radiation patterns.
The derived SCF is validated through Monte Carlo simulations under a wide range of system parameters, including antenna spacing, array geometry, and different azimuth and zenith angular spreads. The analytical results are shown to closely match the simulation outcomes in all considered scenarios, demonstrating the accuracy and applicability of the proposed model for HAPS-based mMIMO systems.

\appendices
\section{Useful Lemma}
\label{app:A}
\begin{lemma}
\label{lemma:lemma1}
In a 3D propagation environment, the antenna array response can be expanded using the spherical harmonic expansion (SHE) of a plane electromagnetic wave given by \cite{colton1998inverse}, 
\begin{equation}
e^{i k \hat{\mathbf{v}} \cdot \mathbf{x}}=\sum_{n=0}^{\infty} i^n(2 n+1) j_n(k\|\mathbf{x}\|) P_n(\hat{\mathbf{v}} \cdot \hat{\mathbf{x}}), \mathbf{x} \in \mathbb{R}^3
\end{equation}
where $j_n$ is the spherical Bessel function of order $n$ and $P_n$ is the Legendre polynomial function of order $n$. Let $(\phi_1,\theta_1)$ and $(\phi_2,\theta_2)$ be the spherical coordinates of vectors $\hat{\mathbf{v}}$ and $\mathbf{x}$ respectively, then by the Legendre addition theorem \cite{kennedy2013hilbert}.
\begin{eqnarray}
\label{Eq: Legendre addition theorem}
    P_n(\hat{\mathbf{v}} \cdot \hat{\mathbf{x}})&=&\frac{4 \pi}{2 n+1} \sum_{m=-n}^{m=n} Y_n^m(\hat{\mathbf{v}}) Y_n^{m *}(\hat{\mathbf{x}}) \nonumber \\
    &=& P_n\left(\cos \theta_1\right) P_n\left(\cos \theta_2\right)+2 \sum_{m=1}^n \frac{(n-m)!}{(n+m)!} \nonumber\\
    &\times& P_n^m\left(\cos \theta_1\right)P_n^m\left(\cos \theta_2\right) \cos \left[m\left(\phi_1-\phi_2\right)\right].
    \nonumber \\
\end{eqnarray}
\par
\end{lemma}

\begin{lemma}
\label{lemma:lemma2}
For non-negative integers $n$ and $m$ \cite{hofsommer1960table},
\begin{equation*}
P_{2 n}(\cos x)=p_n^2+2 \sum_{k=1}^n p_{n-k} p_{n+k} \cos (2 k x),
\end{equation*}
\begin{equation*}
P_{2 n-1}(\cos x)=2 \sum_{k=1}^n p_{n-k} p_{n+k-1} \cos ((2 k-1) x),
\end{equation*}
\begin{equation*}
\bar{P}_{2 n}^{2 m}(\cos x)=\sum_{k=0}^n c_{2 n, 2 k}^{2 m} \cos (2 k x),
\end{equation*}
\begin{equation*}
\bar{P}_{2 n}^{2 m-1}(\cos x)=\sum_{k=1}^n d_{2 n, 2 k}^{2 m-1} \sin (2 k x)
\end{equation*}
\begin{equation*}
\bar{P}_{2 n-1}^{2 m}(\cos x)=\sum_{k=1}^n c_{2 n-1,2 k-1}^{2 m} \cos ((2 k-1) x)
\end{equation*}
\begin{equation*}
\bar{P}_{2 n-1}^{2 m-1}(\cos x)=\sum_{k=1}^n d_{2 n-1,2 k-1}^{2 m-1} \sin ((2 k-1) x),
\end{equation*}
where $p_n, c_{2 n, 2 k}^{2 m}, c_{2 n-1,2 k-1}^{2 m}, d_{2 n, 2 k}^{2 m-1}$ and $d_{2 n-1,2 k-1}^{2 m-1}$are given
by the recursion relations in \cite{hofsommer1960table}.
\end{lemma}
\bibliography{my_bibliography}
\bibliographystyle{IEEEtran}
\end{document}